\documentclass[a4paper,fleqn]{cas-dc}
\usepackage[numbers]{natbib}
\usepackage{subcaption}
\usepackage{graphicx}
\usepackage{flafter}
\usepackage{gensymb} 
\usepackage{ulem}
\usepackage{longtable}
\usepackage{float}
\usepackage{lineno}
\modulolinenumbers[5]

\ExplSyntaxOn
\keys_set:nn { stm / mktitle } { nologo }
\ExplSyntaxOff

\begin{document}
\let\WriteBookmarks\relax
\def\floatpagepagefraction{1}
\def\textpagefraction{.001}
\shorttitle{Stopping power of fission fragments in Mylar and nickel foils}
\shortauthors{Th. Materna et~al.}

\title [mode = title]{Stopping power of fission fragments in thin Mylar and nickel foils}                      

\author[1]{T. Materna}[orcid=0000-0002-7806-3592] 
\cormark[1] 
\ead{thomas.materna@cea.fr}
\author[1]{E. Berthoumieux}
\author[1]{Q. Deshayes}
\author[1]{D. Dor\'e}
\author[1]{M. Kebbiri}
\author[1]{A. Letourneau}
\author[1]{L. Thulliez}
\author[2]{Y.H. Kim}
\author[2]{U. K\"oster}
\author[3]{X. Ledoux}

\address[1]{IRFU, CEA, Universit\'e Paris-Saclay, 91191 Gif-sur-Yvette, France}
\address[2]{Institut Laue-Langevin, 38042 Grenoble Cedex 9, France}
\address[3]{Grand Acc\'el\'erateur National d’Ions Lourds, CEA/DRF - CNRS/IN2P3, 14076 Caen, France}

\cortext[cor1]{Corresponding author}

\begin{abstract}
The energy loss of heavy ions in thin Mylar and nickel foils was measured accurately using fission fragments from $^{239}Pu(n_{th},f)$, mass and energy separated by the Lohengrin separator at ILL. 
The detection setup, placed at the focal plane of the Lohengrin separator enabled to measure precisely the kinetic energy difference of selected fragments after passing through the sample. 
From these data, the stopping powers in Mylar and nickel layers were extracted and compared to calculations. 
Whereas large deviations are observed with SRIM-2013 for Mylar, fairly good agreements are obtained with the semi-empirical approach of Knyazheva {\it et al.} and the calculations contained within the DPASS database. 
In nickel,  SRIM-2013 and Knyazheva model are in agreement with our data within about 10 \%, while large deviations are observed with DPASS. 
We used our data to provide updated parameters for the Knyazheva {\it et al.} model and rescale DPASS database for nickel and Mylar.
\end{abstract}


\maketitle
\normalem
\section{Introduction}
\label{sec:1}
The slowing down phenomena of heavy ions passing \allowbreak through matter is a complex process which is difficult to model accurately. Stopping power calculations are mainly based on the effective charge and the velocity of the ions. For heavy ions, screening effects modify the Coulomb potential. In addition, both the charge and the velocity vary along the path. While for elemental targets the problem is already not simple, it becomes very complex for compound targets. All these aspects and a lack of experimental data probably explain the absence of a predictive stopping power theory although it is requested in different fields (nuclear physics, hadron therapy, dosimetry, ...) 

In neutron-induced fission reactions, emitted fragments have low kinetic energies (0.4 $\lesssim$ E/A $\lesssim$ 1.2 MeV/u for binary fragments) and a wide range of nuclear charges (28 $\lesssim$ Z $\lesssim$ 64). These fragments can lose a large part of their energy in thin material layers.
Accurate energy loss estimates for such heavy ions are needed in the context of the renewed interest for multi-parameter fission experiments. Recently, new spectrometers for neutron-induced fission have been developed \cite{DORE2014,MEIERBACHTOL2015,FREGEAU2016,TSEKHANOVICH2018}. Most of them use emissive foils and/or gaseous detectors to detect fission fragments and measure their velocity and energy. The masses of the fragments are then reconstructed by combining these two quantities. Typical experimental setups include several layers of thin foils leading to significant energy loss which has to be known to be corrected for.

In the Falstaff spectrometer \cite{DORE2014} developed at CEA (Fran\-ce), the velocities of the two fragments are measured thanks to two Mylar emissive foils (0.5 $\mu$m thick) combined with a MWPC (MultiWire Proportional Counter). Ionization chambers made up of Mylar entrance windows (0.9$\mu$m) provide the kinetic energy information. The foreseen actinide targets will be deposited on a nickel backing. It is then mandatory to estimate precisely the energy loss in the different layers in order to provide accurate information on the fission process. 

There is few data on energy loss for slow heavy ions \cite{MONTANARI2017,WITTMAACK2016}. This is especially striking for compound material like Mylar as it is shown in Fig. 3 of \cite{MONTANARI2017}. 
Some specific data on fission fragments exists, e.g. \cite{KNYAZHEVA2006}. They were obtained using a spontaneous fission source and a "2-v, 1-E" spectrometer. However, the limited mass resolution gives rise to beams composed of several masses and nuclear charges. 

In the literature, as detailed later in section \ref{sec:2}, available models do not agree and do not always reproduce experimental data. 
Differences between models could lead to large discrepancies if they are applied to the mass calculation using $A=2E/v^2$. For example, a Cs fragment having a given velocity and detected with an energy of 65 MeV after passing 0.5 $\mu$m of Mylar will have its energy corrected by 3.8 MeV with LISE++ \cite{bazin_program_2002,tarasov_lise_2008} and by 5.4 MeV with Geant4-EMZ \cite {G4EMZ}. It leads finally to a difference of 4 amu in the mass calculation. For light fragments, the discrepancy  leads to differences around 1 amu.

The scarcity of experimental data and the unreliability of models have motivated an experimental campaign at the Lohengrin separator \cite{ARMBRUSTER1976,FIONI1993} of the Institut Laue-Langevin (France) to measure the energy loss of a large number of different fission fragments in thin Mylar and nickel foils. The objective of this campaign was to bring high precision data for well selected fission fragments to be used to improve the modelling of the energy loss in the entire region of interest for fission fragments.
Experimental data are compared with some models available in the literature or transport codes as Geant4.

The paper is organized as follows. In section 2, some energy loss models available in the literature are discussed. 
The  experiment is described in the section 3 and the analysis detailed in section 4. Results and comparison with models are shown in section 5.

\section{Models and Data}
\label{sec:2}

\begin{figure}
	\centering
	\includegraphics[width=\linewidth]{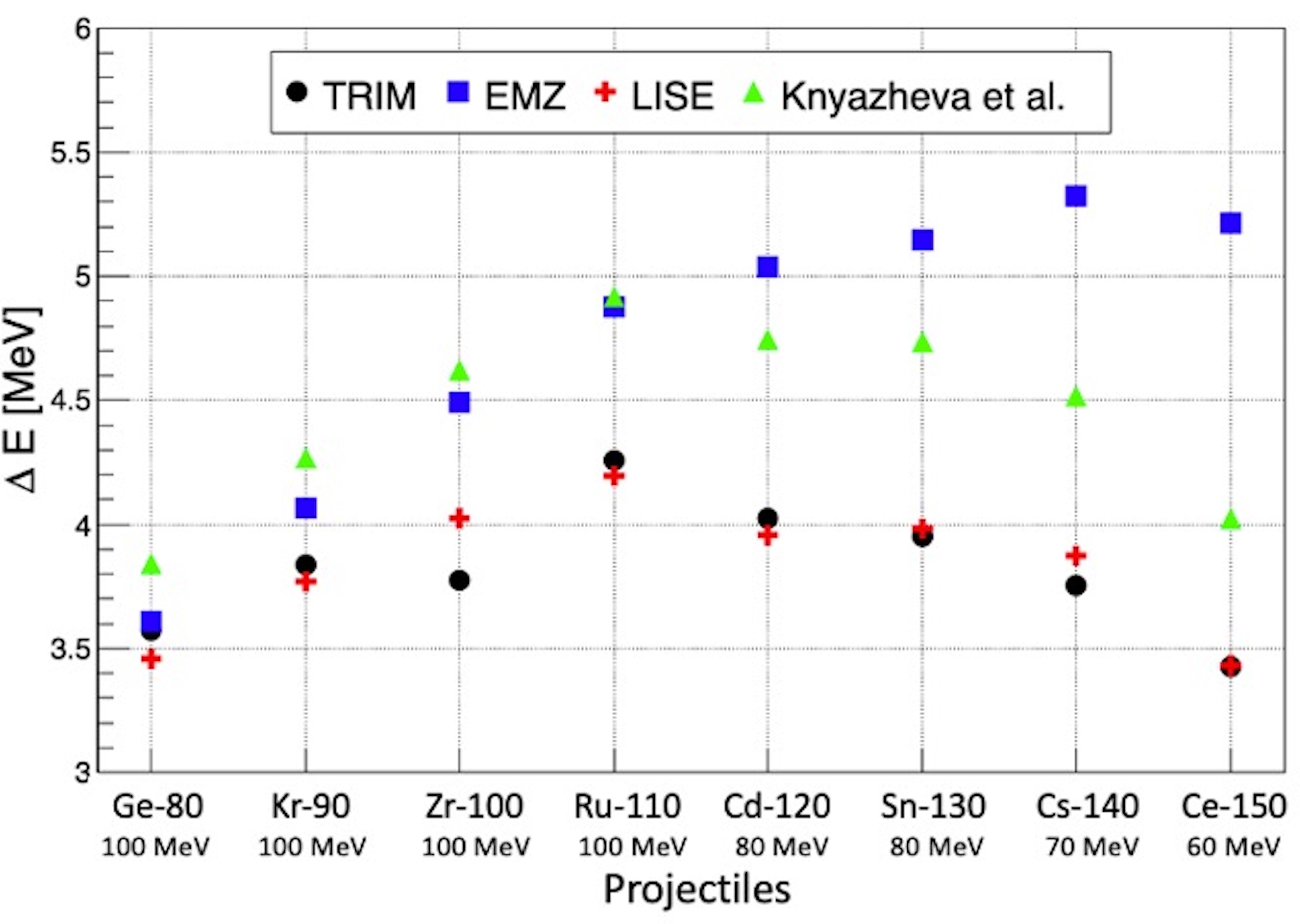}
	\caption{Energy loss of various fission fragments ($^{80}$Ge, $^{90}$Kr, $^{100}$Zr, and $^{100}$Ru at 100 MeV; $^{120}$Cd and $^{130}$Sn at 80 MeV; $^{140}$Cs at 70 MeV, $^{150}$Ce at 60 MeV) in a 0.5 $\mu$m thin Mylar foil according to SRIM-2013 \cite{ziegler_srim_2010} (denoted TRIM), GEANT-4.10 EMZ \cite{G4EMZ}, LISE++ \cite{bazin_program_2002,tarasov_lise_2008} and Knyazheva et al. \cite{KNYAZHEVA2006} prescription.}
	\label{fig:trimg4}
\end{figure}

Contrary to protons and alphas, where energy loss in materials is known in a wide energy range with a good precision, data on energy loss of heavy ions still suffer from an important lack of precision. As an illustration, Fig. \ref{fig:trimg4} shows the energy loss of some fission fragments in a 0.5-$\mu$m Mylar foil calculated with different codes (see figure for details). One observes important discrepancies between the various codes going from 15\% for light fragments up to almost 50\% for heavy ions.
The main reason for these discrepancies is related to the increase in complexity needed to describe the penetration of ions in matter with the increasing atomic number of the incoming ion \cite{sigmund_progress_2016}. 

From a qualitative point of view, one may distinguish three main regimes to describe the energy loss of heavy ions depending on their atomic number, $Z_1$ and speed $v$.
At low speed, \emph{i.e.} below the Bohr velocity $v_0=\alpha\,c$, where $\alpha$ denotes the fine-structure constant and $c$ the speed of light, the electronic stopping power is expected to be proportional to the speed of the incoming ion as depicted by the LSS theory \cite{Lindhard1963}.
In the opposite direction, for relativistic velocities,  i.e for $ 2 Z_1 v_0/v \lesssim 1$ according to the Bohr criterion, the Born approximation is valid, and so the Bethe-Bloch formalism \cite{bethe_1930,bloch_1933} is well suited to describe the stopping power.
In the intermediate region, corresponding to the region of interest for fission-fragment energy loss, a good starting point is the Bohr formalism \cite{sigmund_progress_2016,bohr_velocity-range_1941} based on the classical orbitals. In this energy domain, close or below the Bragg peak, one also has to consider the screening effect that is expected to be significant below the Thomas-Fermi velocity, $v_{TF} = v_0 Z_1^{2/3}$ \cite{sigmund_progress_2016}. 

Based on these ideas, Knyazheva \emph{et al.} \cite{KNYAZHEVA2006} have developed a phenomenological model devoted to describe fission fragment stopping power in some specific materials. The model is based on a polynomial expansion of the Bohr stopping number, $L_{Bohr}$ and an equilibrium ion charge state,  $\overline{Z_1}$, derived from the standard Thomas-Fermi charge:
\begin{eqnarray}
	\overline{Z_1} & = & Z_1 \left[ 1 - \exp{\frac{-v}{v_0 Z_1^\lambda}}\right]
	\label{eq:Zeff}
\end{eqnarray}

where $\lambda $ is an adjustable parameter ($\lambda=2/3$ for the standard Thomas-Fermi case).  The parameters of these models have been tuned to reproduce the energy loss of fission fragments emitted in spontaneous fission of $^{252}$Cf and detected with a "2-v, 1-E" spectrometer giving a mass resolution slightly better than 2~a.m.u. and a $\pm$2.5~MeV energy resolution. This allows to have a very precise modelling of the stopping power but limited to a narrow energy region, for selected materials and for incident ions ranging typically from $34\lesssim Z_1\lesssim 64$.

Also starting from the Bohr classical model or its low energy extension \cite{sigmund_low-speed_1996}, Sigmund and Schinner \cite{sigmund_binary_2002} have developed the ``Binary theory of electronic stopping''. This theory was used for the ICRU73 report \cite{noauthor_stopping_2005} for incident ions up to Ar. More recently calculation were performed for ion up to uranium on any elemental target ranging from hydrogen to uranium \cite{schinner_expanded_2019} and provide as the freely available DPASS database \cite{dpass}.  
The database can cover energies per nucleon ranging from 1~keV/u to 1~GeV/u  in a fully consistent manner, and so is at the state of the art in the treatment of all phenomena to consider to cover the broad range both in Z (of the incident ion and of the target) and energy. In our region of interest and in the case of compound materials the Bragg rule can be applied since valence structure effects are strongly reduced when screened ions are employed as in case of fission fragments \cite{sigmund_valence_2005}.

Few programs allow to compute energy loss of any ion in any material in a broad energy range \cite{natanasabapathi_stopping_2012}. The most popular program nowadays is certainly SRIM-2013 \cite{ziegler_srim_2010}. SRIM is initially based on the Bethe-Bloch formalism and suited for high energy. It was extended in a semi phenomenological way along the years to cover all possible combinations, but leads sometimes to some inconsistencies \cite{wittmaack_misconceptions_2016}.
To account for charge screening SRIM is relying on the Brandt and Kitagawa theory \cite{brandt_effective_1982,ziegler_stopping_1988}.

Other programs are available to estimate energy loss for a wide range of ion-target combinations. The two other codes considered in this study are LISE++ \cite{bazin_program_2002,tarasov_lise_2008} and ATIMA \cite{ATIMA}. 
Nevertheless below 10 MeV/u, our region of interest, both codes are based on a former version of SRIM 
\cite{ziegler_stopping_1985,ziegler_stopping_1999}, and then no significant deviation with the actual SRIM version is expected, as shown in Fig \ref{fig:trimg4}.

Monte-Carlo simulation is a key ingredient in many nuclear experiments. For that purpose, one commonly uses the Geant4 toolkit \cite{G4EMZ}. In our region of interest (ion of energy below 2 MeV/u) two main models are available for the ion energy loss. The ``default'' model called \emph{G4BraggIonModel} is based on He stopping given by the ICRU49 \cite{berger_report_1993} report, scaled to any ion by using the effective charge concept developed by Ziegler and Manoyan \cite{ziegler_stopping_1988} based on the Brandt and Kitagawa theory \cite{brandt_effective_1982}. An alternate model called \emph{G4Ion\-Parametrised\-LossModel}
\cite{lechner_validation_2010} is based on the ICRU73 report \cite{noauthor_stopping_2005} for incident ions up to Ar. For ions heavier than Ar, as in fission fragments, a scaling is applied to the Ar stopping power based on the standard Thomas-Fermi charge. The scaling factor is given by:
\begin{equation}
	\left( \frac{Z_1}{Z_{Ar}} \frac{1 - \exp{\frac{-v}{v_0 Z_1^{2/3}}}}{1 - \exp{\frac{-v}{v_0 Z_{Ar}^{2/3}}}} \right)^2
	\label{eq:scalicru73}
\end{equation}
with $Z_{Ar} = 18$. Results of this scaling applied to specific fission fragments are also shown in Fig. \ref{fig:trimg4} with the label \emph{EMZ}.

\begin{figure}[pos=htb]
	\centering
	\includegraphics[width=\linewidth]{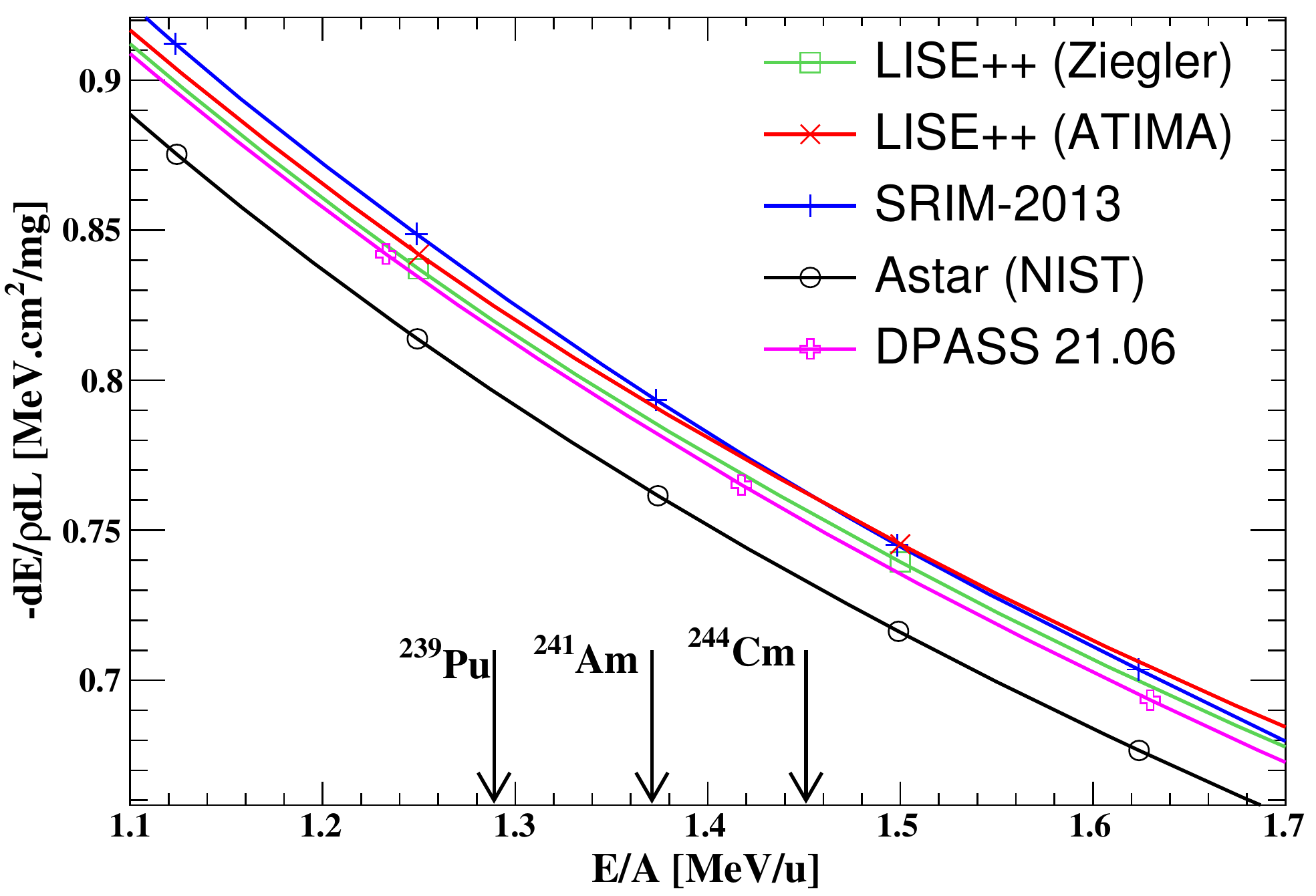}
	\caption{$\alpha$ stopping power in Mylar around 1.5~MeV/u given by various codes (LISE++ based on SRIM \cite{tarasov_lise_2008}, LISE++ based on ATIMA \cite{ATIMA},  SRIM-2013 \cite{ziegler_srim_2010}) and tables (DPASS \cite{dpass} based on PASS \cite{schinner_expanded_2019}, Astar \cite{astar_nist} based on ICRU49 \cite{berger_report_1993}). Arrows indicate the main energy peak of the triple-$\alpha$ source used in this study.}
	\label{fig:alphadEdX}
\end{figure}

Finally since the determination of our sample thickness for this experiment is crucial and based on $\alpha$ energy loss, we made a comparison of various codes and tables around 1.5 MeV/u. Results are shown in Fig. \ref{fig:alphadEdX} together with main energy peak positions of the triple-$\alpha$ source used to determine the sample thicknesses. For Mylar samples, most of the codes and tables are in agreement within 1\% in this region except the Astar table \cite{astar_nist} (aka ICRU49 \cite{berger_report_1993}) that lies about 3\% below the other codes/tables.
The thickness determination shown hereafter was performed using the SRIM-2013 \cite{ziegler_srim_2010} code.

\section{Experiment}
\label{sec:3}
The experiment was performed at the high-flux reactor of Institut Laue Langevin (ILL) using the Lohengrin separator to select fission fragments according to their mass, ionic charge and kinetic energy \cite{ILLDATA}. The general procedure was to measure the kinetic energies of selected fission fragments with and without thin foils of Mylar and nickel placed at the exit of Lohengrin. Moreover, samples were characterized separately, at CEA Saclay, using a triple alpha source and a dedicated set-up. The experimental details are given in the following.

\subsection{Sample thickness measurement}
\label{sec:sample_thickness}
\begin{figure}[pos=htb]
	\centering
	\includegraphics[width=\linewidth]{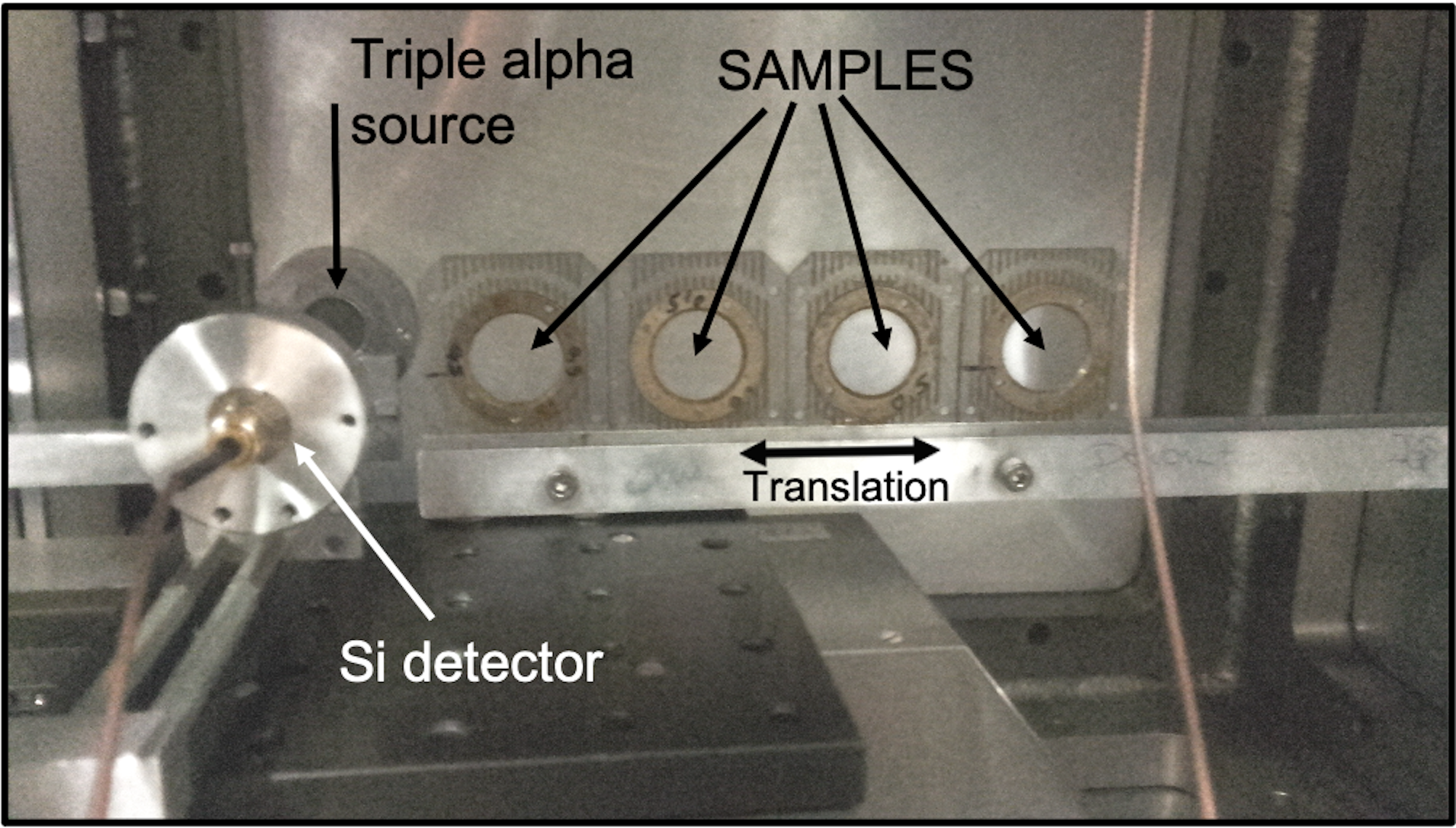}	\caption{Sample characterisation setup. Four samples are moved horizontally between the triple alpha source and the PIPS detector. }
	\label{fig:alphaSetUp}
\end{figure}

Measurements of sample thicknesses were performed using the alpha transmission method with a triple alpha source of $^{239}$Pu, $^{241}$Am and $^{244}$Cm. Kinetic energies of alpha particles were measured with and without sample using a PIPS detector having an energy resolution of 15 keV and placed behind the sample (Fig. \ref{fig:alphaSetUp}). Samples were placed on a trolley which moved perpendicularly to the source-detector axis allowing to place correctly the sample and to measure thickness variations from one end of the sample to the other. The reaction chamber was kept at a pressure below $10^{-5}$ mbar. 

Different positions without samples (before, between and after the samples) were used to calibrate the detector. It has been checked that results are similar for different points on a sample. Then the statistics was summed over all points. 

\begin{figure}[pos=htb]
	\centering
	\includegraphics[width=\linewidth]{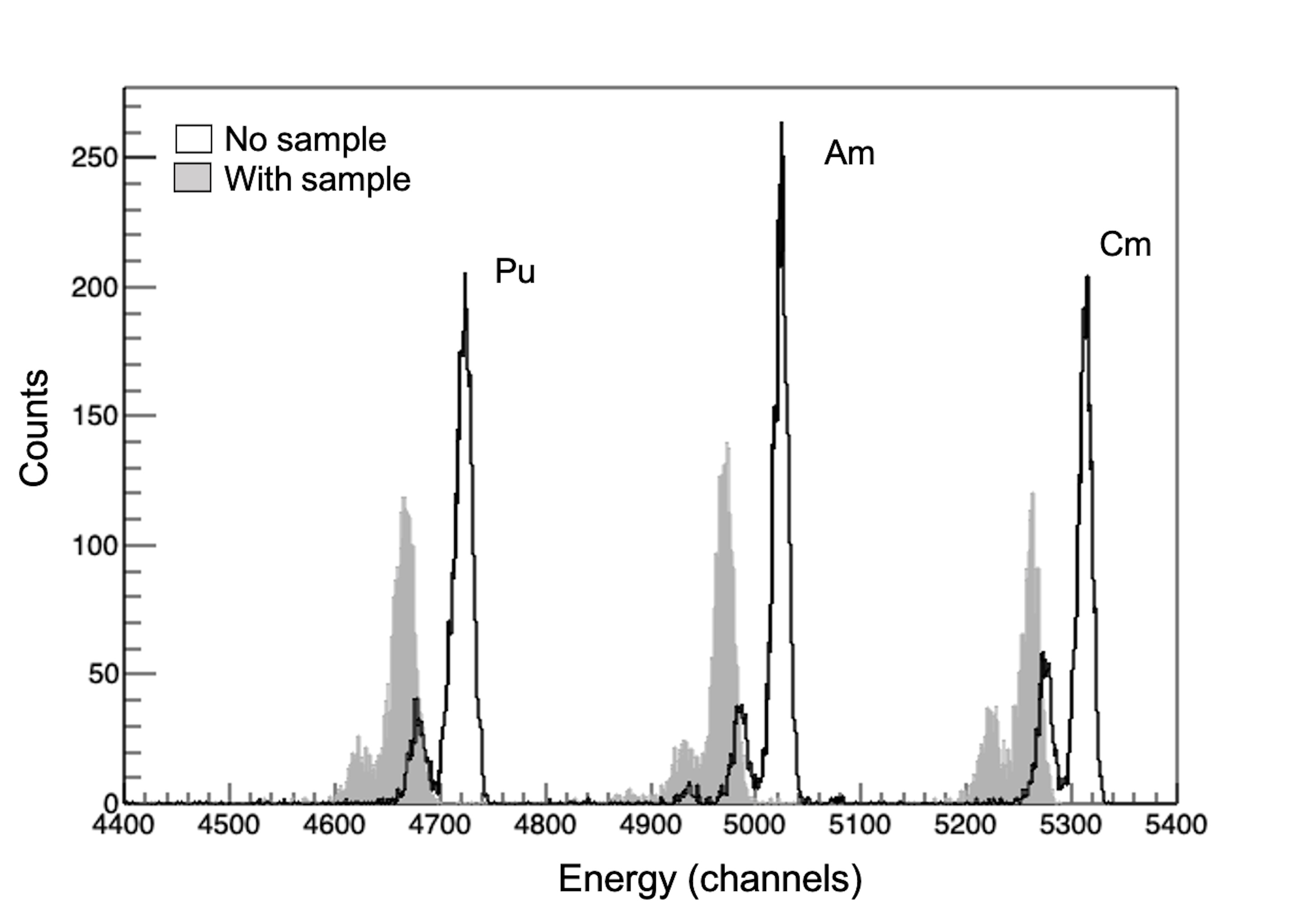}	\caption{Alpha energy distributions without sample (empty histograms) and with sample (grey histograms).}
	\label{fig:spectres}
\end{figure}

Figure \ref{fig:spectres} shows a spectrum obtained without sample (emp\-ty histogram) and a spectrum associated to the 0.5 $\mu$m-thick Mylar sample (grey histogram). The two main contributions for each actinide are clearly seen. 
All distributions without samples were fitted separately with Gaussian functions to obtain the mean channel value of each peak. A linear energy calibration of the detector was obtained by fitting the energy values of the alphas for $^{239}$Pu, $^{241}$Am and $^{244}$Cm (from \cite{NNDC_Cm,NNDC_Am,NNDC_Pu}) as a function of these channel values. In the same way, distributions obtained with Mylar foils were also fitted with Gaussian functions and converted into energy using the energy calibration. Only the main peaks for $^{241}$Am and $^{244}$Cm were considered.

The energy difference between results with and without samples was then calculated and associated to a Mylar (nickel) thickness via SRIM calculations using a density value of \mbox{1.397 g/cm${^3}$} (\mbox{8.896 g/cm${^3}$}).  An uncertainty of 0.4\% was added quadratically to the obtained thickness uncertainty to account for a possible misalignment of the foil in the setup by $\pm$5\degree. In Table \ref{table:samplenm} averaged thicknesses calculated with the $^{241}$Am and $^{244}$Cm main contributions are listed with their uncertainty value for the three samples. Results will be used for the stopping power calculations in section 4.5. 

\begin{table}[pos=htb]
	\centering
	\caption{Measured averaged thicknesses for Mylar and nickel samples.}
	\begin{tabular*}{\columnwidth}{c c c}
		\hline
		\textbf{Samples} & \textbf{<Thickness> [nm]}  & \textbf{Uncertainty [nm]} \\
		\hline
		A (Mylar) & 520.7 & 2.7 \\
		B (Mylar) & 753.6 & 3.7 \\
		C (Nickel) & 259.1 & 1.2 \\
		\hline
	\end{tabular*}
	\label{table:samplenm}
\end{table}

\subsection{Ion selection with Lohengrin}
\label{sec:3.1}

Lohengrin is a mass separator with a direct view to an actinide target placed close to the core of the reactor, where the flux of thermal neutrons is about $5\;10^{14}\; n\;s^{-1} cm^{-2}$ \cite{ARMBRUSTER1976}. 
Here, a highly enriched $^{239}$Pu target (300 $\micro$g/cm$^2$ and 70 x 7 mm$^2$ large) was used and it was covered with a 0.25 $\micro$m-thick Ni foil to reduce sputtering losses \cite{Koster2010}. At the time of the experiment this target had already been exposed to the neutron flux for over five weeks, which leads to some redistribution of target material by sputtering and thermal diffusion into the backing and cover foil \cite{Koster2010} as well as laterally. The additional energy loss of fission fragments emitted from deeper layers of the target backing leads to kinetic energies that are unusually low for fission fragments. For the present experiment such a broadening of the natural kinetic energy distribution of fission fragments is beneficial as it allows selecting ions over a wider energy range with the Lohengrin spectrometer. 

Neutron induced fission of $^{239}$Pu produces mostly two fragments that are emitted in opposite directions, a heavy fragment with a mass number around 140 and a light fragment with mass around 100. The energy of the light fragment peaks around 105 MeV while the energy of the heavy fragment depends more strongly on its mass and ranges between about 50 and 80 MeV \cite{Plompen2020}. At Lohengrin, the combination of magnetic and electrostatic fields allows the selection of the fragments according to their mass over ionic charge (A/q) and to their kinetic energy over ionic charge (E/q) ratios. 
The ionic charge of the fragments mainly results from the passage of the fragments in the Ni foil that covers the actinide target. The ionic-charge distribution is therefore usually about Gaussian and ranges from about 17+ to 25+. For some fragments, however, the distribution is strongly deformed and can reach 30+. It is due to the presence of nanosecond isomeric states in the de-excitation cascade that decay in the path between the target and the spectrometer by the ejection of an electron and finally lead to the emission of several electrons in the reorganisation of the atomic shell \cite{SCHMITT1984, Materna2009a}. Only few nuclides have such nanosecond isomeric states and setting the spectrometer to a particularly high ionic charge state allows to produce a quasi isotopically pure beam.

\begin{figure}[pos=htb]
	\centering
	\includegraphics[width=1.0\linewidth]{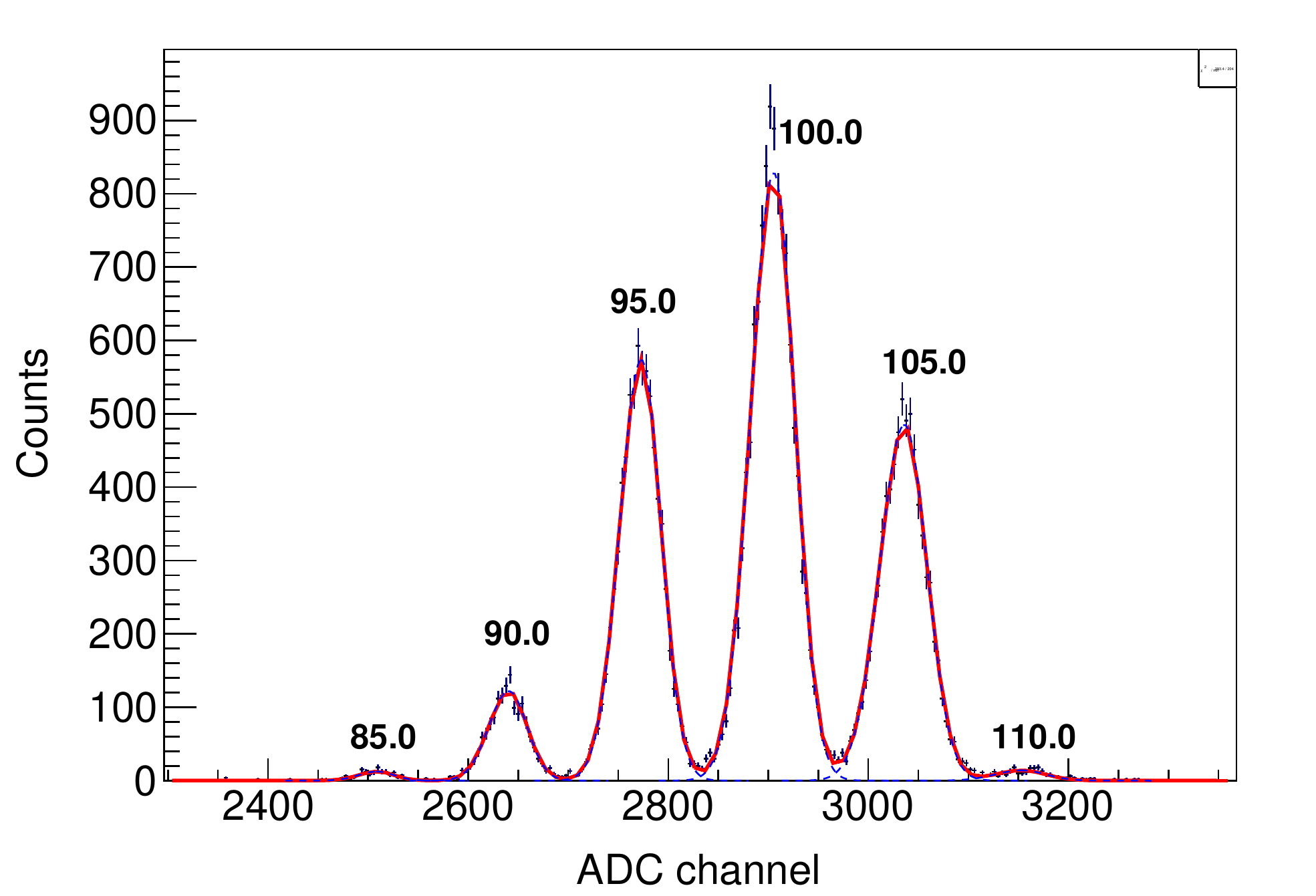}
	\caption{Measured spectrum for Lohengrin parameters set to A/q = 100/20 and E/Q = 100/20. The beam is composed of masses 85, 90, 95, 100, 105 and 110. See text for details.}
	\label{fig:fig10020100}
\end{figure}

\begin{figure}[pos=htb]
	\centering
	\includegraphics[width=1.0\linewidth]{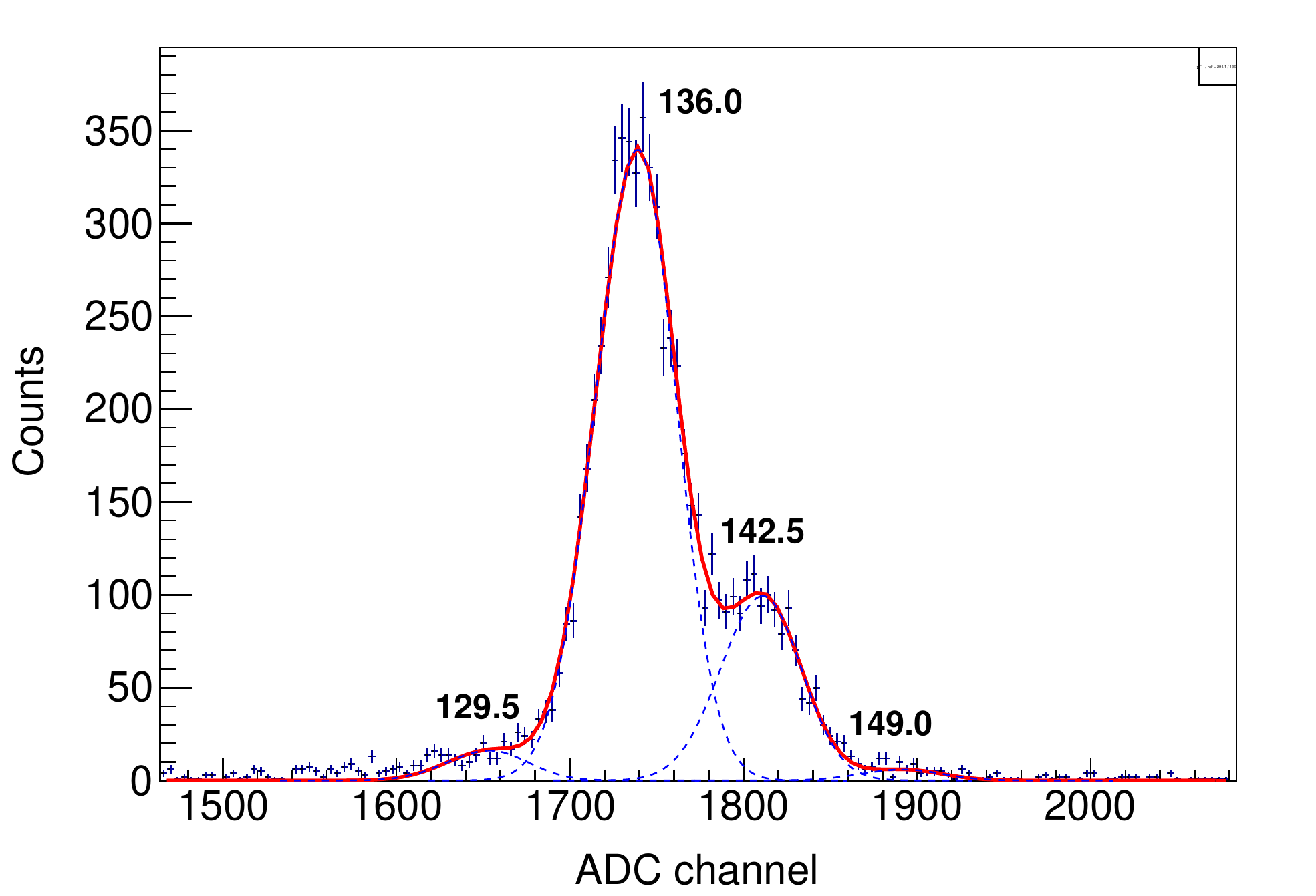}
	\caption{Measured spectrum for Lohengrin parameters set to A/q = 136/21 and E/q = 65/21. The beam is composed of masses 136 and 149 and contaminated by mass 129,130 and 142,143.}
	\label{fig:spectre1362165}
\end{figure}

The beam at the exit of Lohengrin is often composed of several masses; they correspond to the different possible ionic charges leading to the same or very similar A/q ratio and they can be distinguished by their kinetic energy. 
Figure \ref{fig:fig10020100} shows the spectrum measured in a PIPS detector at the exit of Lohengrin when the Lohengrin parameters were set to A/q = 100/20 and E/q = 100/20 MeV. The beam is mainly composed of 6 masses (85, 90, 95, 100, 105 and 110) corresponding respectively to the ionic charge (17, 18, 19, 20, 21 and 22) and energy (85, 90, 95, 100, 105 and 110) MeV. All 6 masses can be used to study the energy loss in a single measurement. Another example is shown in Fig. \ref{fig:spectre1362165} with Lohengrin parameters set to A/q = 136/21 and E/q = 65/21 MeV. In this case, only mass 136 is useful (A/q = 136/21 = 6 is an integer). The other visible peaks (with non-integer ratio for A/q) correspond to fragments emitted from the laterally extended fission source (due to sputtering, cf. above) that exit the spectrometer slightly off-axis. The collimator is large enough to let a residual part of them reaching the detector. Their energy is not well defined and we do not use them for energy loss measurements. However, these contamination peaks must be identified and fitted in the analysis in order to evaluate correctly the position of the main peaks, which are used for the energy loss measurement.


Mass and energy resolutions of the separator depends on the target size, the exit collimator and the detector size. In the energy-loss campaign configuration, the mass resolution is estimated to about 0.5\% and the energy resolution to about 1\%.
The uncertainty on the (mean) energy of the beam can be decomposed into a systematic uncertainty (linked to the absolute energy calibration), expected to be better than 0.2\% and a reproducibility uncertainty, that we estimated to about 0.01\%. 

Since there is no separation in atomic number (Z), a mass peak is composed of different nuclei. The exact beam composition depends on the fission yields and usually, the fission process produces two or three major Z per mass. As explained above, it is however possible, in some rare cases, to produce a quasi pure beam thanks to the selection of a high ionic charge. This was applied to obtain a beam of $^{140}$Cs using Lohengrin parameters set to A/q = 140/30 and $^{144}$Cs using A/q = 144/27.     

\subsection{Energy loss measurement set-up}
\label{sec:3.2}

\begin{figure}[pos=htb]
	\centering
	\includegraphics[width=1.0\linewidth]{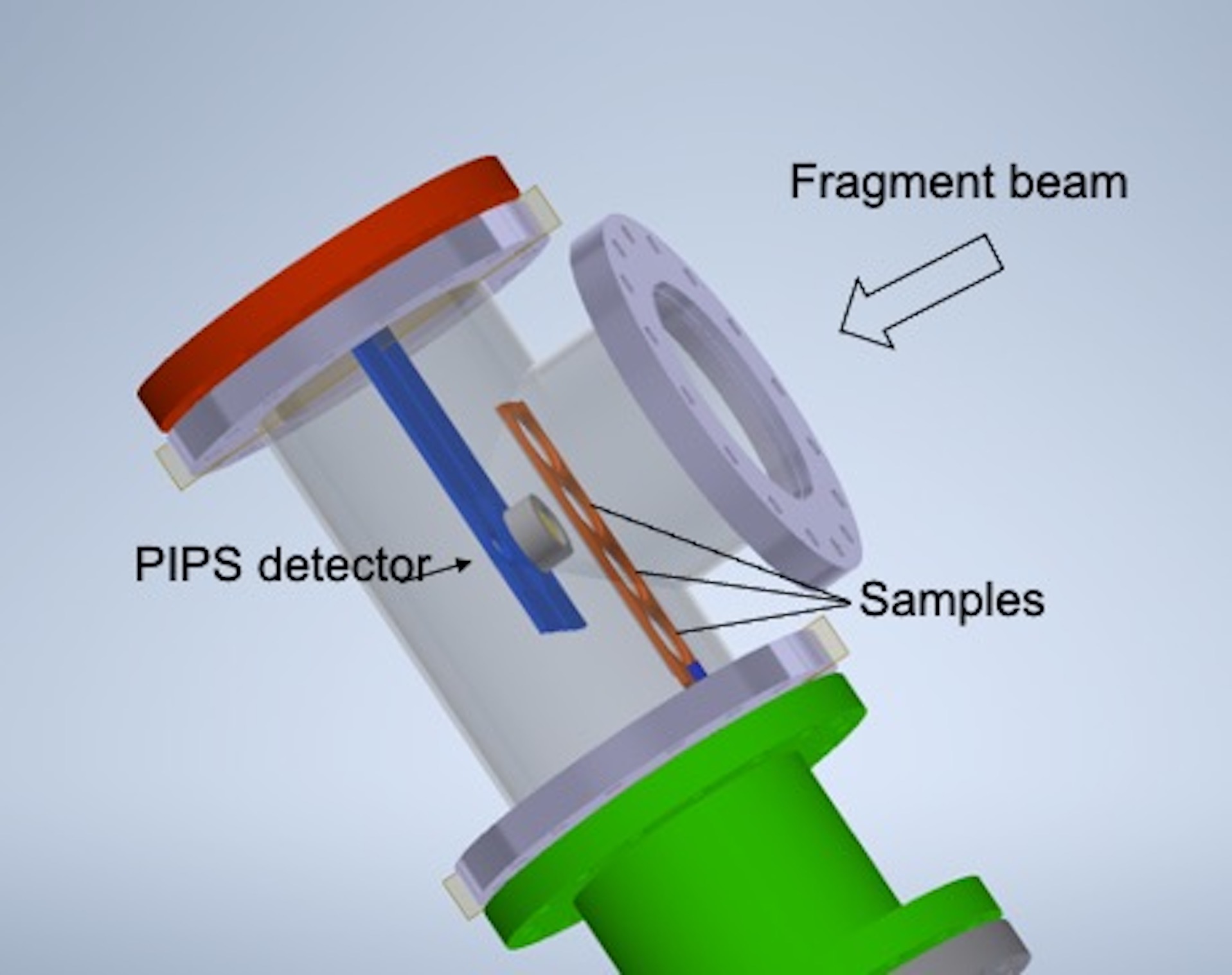}
	\caption{Schematic drawing of the sample-detection system.}
	\label{fig:newsetup_col_300dpi.jpg}
\end{figure}

The detection system for the energy loss measurement is similar to the system described in the sample characterization section. A target holder with three samples and a hole (reference for calibration) moved perpendicularly to the fragment beam (Fig. \ref{fig:newsetup_col_300dpi.jpg}). A PIPS detector was placed behind the target holder. The setup was installed at the straight exit of the LOHENGRIN spectrometer. The sample and the PIPS detector were placed slightly off-center, by about 16 mm, of the spectrometer exit. It results in a shift of the mean energy of the fragments on the detector by about -0.175 \% that was taken into account in the analysis process. 

\subsection{Experimental Procedure}
\label{sec:3.3}

The experimental procedure consisted of alternative measurements without and with the sample in the beam for a given A/q setting of the spectrometer in order to obtain a precise calibration of the detector before measuring the energy losses in the sample. 

{\small
	\begin{table}[pos=ht]
		\centering
		\caption{Lohengrin settings used for the experiment. Samples A, B and C referred to Table 1.}
		\begin{tabular*}{\columnwidth}{ m {2.5cm} m {1.5cm} m {1.5cm} m {1.1cm}  } 
			\hline
			Masses & Energies (MeV) & Ionic charges & Samples \\ 
			\hline
			100 & 70-107 & 20 & A,B,C \\
			130-133 & 65-75 & 20 & A \\
			134-135, 137-139 & 65-75 & 21 & A \\
			136 & 60-75 & 21 & A,B,C \\
			140 & 50-85 & 20 & A,B,C \\
			140 & 55-75 & 30 & A,B \\
			144 & 50-65 & 22 & A,B \\
			144 & 50-65 & 27 & A \\
			141-143,145-147 & 65-75 & 22 & A \\
			149 & 45-60 & 20 & C\\
			100 & 90,100 & 19-26 & A \\ 
			136 & 70 & 20-26 & A \\
			\hline
		\end{tabular*}
		\label{tab:settings}
	\end{table}
}

For light fragments the Lohengrin setting A/q=100/20 allowed measuring energy losses for six masses (85, 90, 95, 100, 105 and 110) at once as explained in subsection \ref{sec:3.1}. For heavy fragments, the ionic charge was selected in order to optimize the desired mass intensity (or signal-to-noise) in the beam with respect to contaminants. In addition, for mass 140 and 144 we performed energy loss measurements at a large ionic charge to obtain pure beams of $^{140}$Cs and $^{144}$Cs respectively. Table \ref{tab:settings} details Lohengrin settings (masses, energies and ionic charges) used for each sample.

In addition to the energy scans described above, a special attention has been paid to the effect of the ionic charge on the energy loss for both light and heavy fragments. We measured the energy losses in Mylar for A=100 at E=90 MeV and at E=100 MeV, with the ionic charge scanned between 19 and 26. The same study was then performed in Mylar for A=136 at E=70 MeV with the ionic charge scanned between 20 and 26. 

To avoid too frequent sample changes, which in our setup had to be achieved by moving manually the target holder in front of the beam, the experimental procedure was finally adapted to calibrate the detector for several masses with no sample in the beam before measuring the energy losses for them with a sample in the beam. The energy stability of the detector was monitored by measuring the peak position for some reference masses (mainly mass 136) periodically.

\section{Data analysis}
\label{sec:5}
The large data set motivated the development of a software called PARZIVAL to handle the whole analysis process from the automatic fitting of all the spectra to the calculation of stopping powers. The first subsection explains our methodology for the calculation of the energy loss. The second one describes detector instabilities and explains their handling. The third and fourth ones detail the analysis process related to detector calibration and energy loss calculation. The fifth subsection explains our method to obtain more precisely stopping powers from energy losses.    

\subsection{Methodology}
\label{sec:5.1}

It is well known that the energy calibration of PIPS detectors varies with the mass of detected ions \cite{Schmitt1965}.  The main idea in this experiment was therefore to alternate, for every mass, an energy calibration of the detector with the sample out of the beam, and energy loss measurements for with the sample in the beam, each time for a small range of energies. 

For a given mass $A$, the calibration of the detector is expected to be linear in energy. The calibration analysis, detailed in subsection \ref{sec:Detec_calib}, consisted in fitting the peak positions ($X$) observed in the detector spectrum for few beam energies defined by the Lohengrin spectrometer ($ E_L$). For all measured masses, such local calibration was found compatible with a linear interpolation in the small range of measured energies: 
\begin{equation} \label{eq:2}
	X  = \alpha_A  +  \beta_A *  E_L
\end{equation}
and indeed both parameters $\alpha_A$ and $\beta_A$ were found to vary significantly with mass. 

Energy losses in a sample were then measured at the same or another set of Lohengrin energies $ E_L'$.
The residual beam energy after the foil, $E_R$, is measured at peak position $X_R$ in the detector spectrum. The energy calibration of the PIPS detector is expected to be still valid :
\begin{equation} \label{eq:2R}
	X_R  = \alpha_A  +  \beta_A *  E_R
\end{equation}
The energy loss is thus calculated from :
\begin{align} \label{eq:3}
	\Delta E &= E_L' - E_R \notag \\ 
	&= E_L' - \frac{X_R - \alpha_A } {\beta_A} 
\end{align}

The uncertainty on the energy loss, $\delta\Delta E$ is calculated from:
\begin{align}
	(\delta\Delta E)^2 =  & (\delta E_L')^2 + (\frac{\delta X_R }{\beta_A})^2 
	+ (\frac{\delta \alpha_A}{\beta_A})^2  \notag \\
	&+ \big( (X_R -\alpha_A) \; \frac{ \delta \beta_A}{\beta_A^2}\big) ^2 \\ 
	&+  2 \; \frac{X_R -\alpha_A}{ \beta_A^3} \; \text{cov}(\alpha_A,\beta_A) \notag
\end{align}
where $ \delta X_R$ is the uncertainty on the peak position, $\delta \alpha_A$ and $\delta \beta_B$ the uncertainties on the two calibration parameters and $\text{cov}(\alpha_A,\beta_A) $ the covariance between the two calibration parameters. 
$\delta E_L'$ denotes the reproducibility part of the uncertainty on the beam energy. Indeed, any global systematic uncertainty on the calibration of the Lohengrin spectrometer would affect in the same way the calibration and the measurement with sample and would cancel in Eq. \ref{eq:3}. The reproducibility part, $\delta E_L'$, already quoted to about 0.01\% of $E_L'$ in subsection \ref{sec:3.1}, is negligible compared to the other uncertainties. 

\subsection{Detector stability}
\label{sec:Detec_stab}

The detector response was monitored in order to detect and correct for possible drifts in detector gain with time. Indeed, as explained above, several hours could separate the measurement of the energy loss in a sample at a specific mass and energy, from the detector calibration at this mass with no sample in the beam. 

Within an hour, no significant drift could be identified above statistical uncertainties. 
For larger periods, a drift in the peak positions was observed, compatible with a detector gain drift. 
Indeed, we found that the position of all the peaks at some measurement time could be related to their position at another measurement time by a unique parameter that does not depend on the associated peak energy: $ X_2 = R(t_2,t_1) X_1 $. 

\begin{figure}[pos=h]
	\centering
	\includegraphics[width=1.0\linewidth]{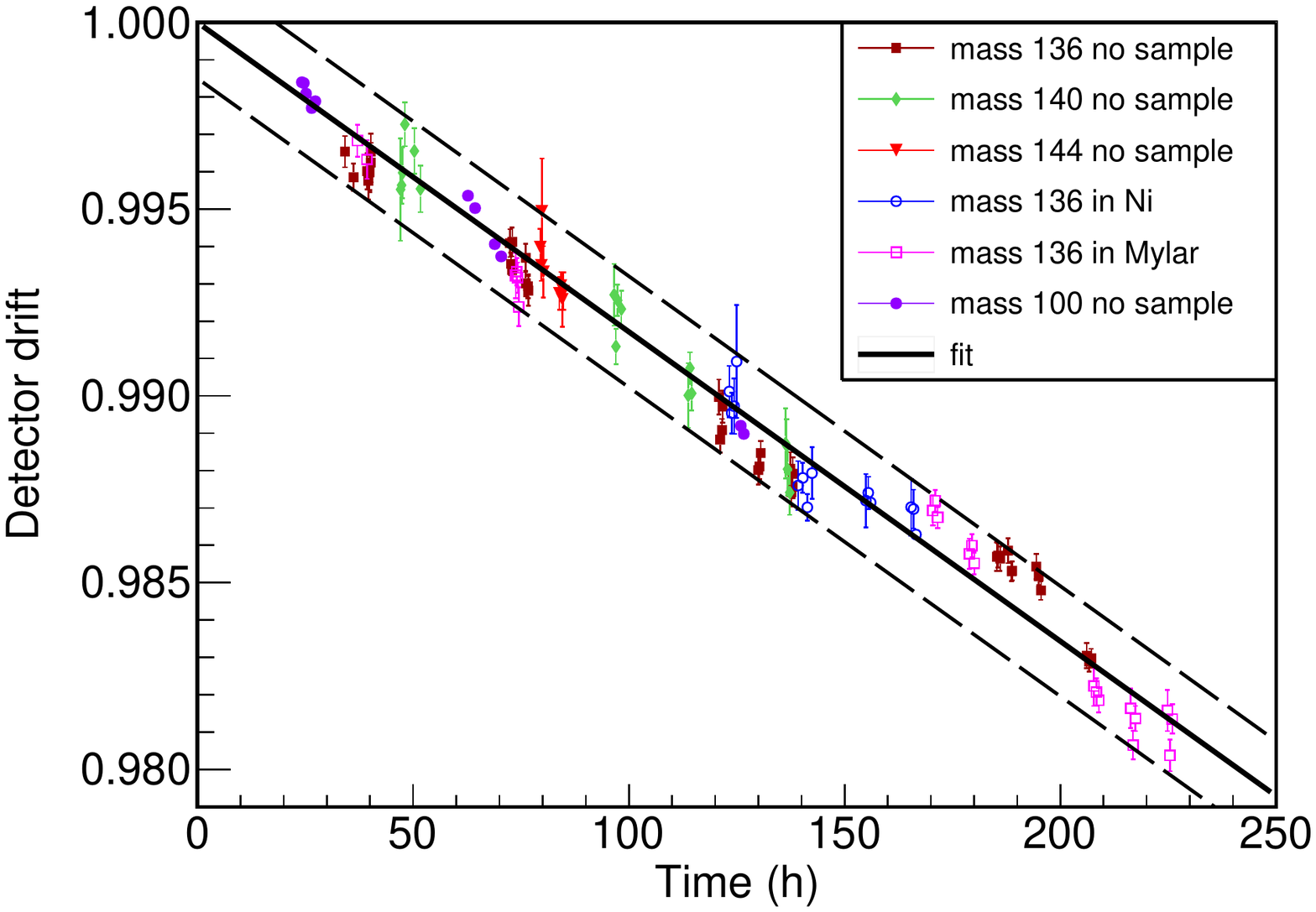}
	\caption{Gain drift of the PIPS detector during the experimental campaign. See text for details.}
	\label{fig:drift}
\end{figure}

The evolution of the gain drift $R(t)$ during the whole measurement campaign, with respect to the start of the campaign, is plotted in Fig. \ref{fig:drift}. The drift is rather linear with time and the evolution can approximately be fitted by $R(t) = 1 - \lambda\: t $ with $\lambda = 8.27 (20) \: 10^{-5} h^{-1}$. The fit is not very good ($\chi ^2/\mbox{NDF} $ = 5) and one can see some data around t = 190 h that differ up to 0.15 \% from the fit, which cannot be explained by their uncertainties. In order to account for this imperfect description of the drift, we added a 0.15\% relative uncertainty to the corrected peak position. This systematic uncertainty is plotted as dashed lines on Fig. \ref{fig:drift}.

\subsection{Detector energy calibration}
\label{sec:Detec_calib}

\begin{figure}[pos=h]
	\centering
	\includegraphics[width=1.0\linewidth]{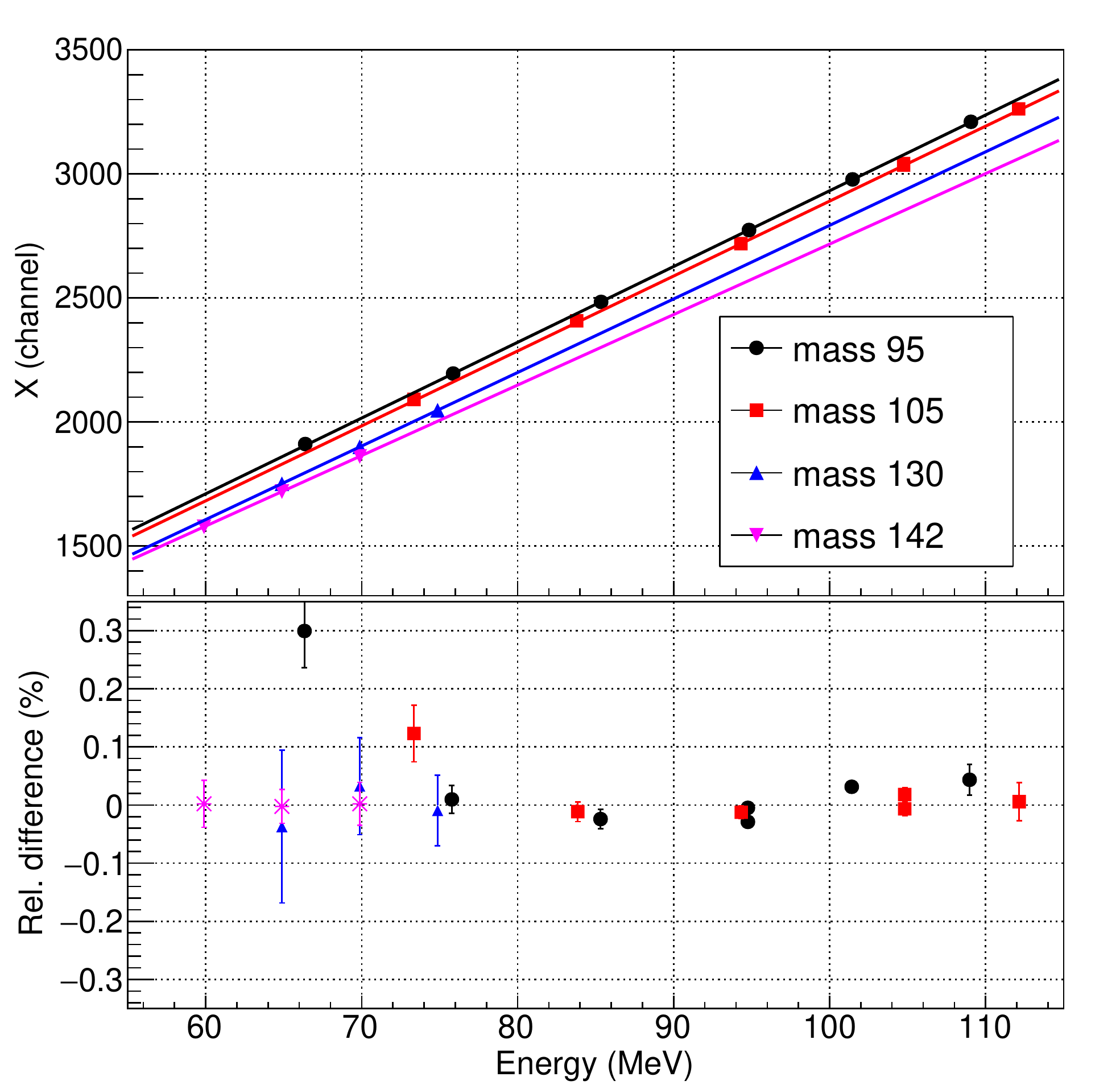}
	\caption{Top: calibration data of the PIPS detector for mass 95, 105, 130 and 142. Bottom:  relative difference between calibration data and their fitted values, $(X-X_{fit})/X_{fit}$, in percents.}
	\label{fig:calibration}
\end{figure}

In principle, identification of the different peaks in the calibration spectra is simple because the beam is composed of few masses defined by the spectrometer parameter. Their intensity can approximately be estimated from fission yields and the detector calibration changes slightly with mass. On the other hand, for an automatic fit processing, some initial values of the position peaks are mandatory to fit all the peaks in a spectrum, especially when they are very close like in Fig. \ref{fig:spectre1362165}. It was solved by using a global parameterization, mass and energy dependent calibration, inspired from the one proposed by Schmitt \emph{et al.} (\cite{Schmitt1965}).
\begin{equation}
	X = (1-c \; A)  \; \alpha_0 +  (1-d \;A) \; \beta_0 \; E_L	
\end{equation}
Parameters of this global model were fitted in an iterative way on a large set of calibration data. The predictability of that parametrization is accurate to about 0.2\% and we use it to give an initial value for all peak positions. Correction for the detector drift is applied. Peaks are modeled as asymmetric Gaussians.  
A precise fit of the local calibration parameters (Eq. \ref{eq:2}) is then performed for each specific mass. Figure \ref{fig:calibration} shows the calibration results for mass 95, 105, 130 and 142. Local calibrations are usually accurate within 0.1\% as shown on the bottom part of the figure. To account for fits with $\chi^2/ \mbox{NDF} $ larger than one, we applied the usual technique consisting in multiplying the fitted parameter uncertainties and covariances by the square root of the $\chi^2/\mbox{NDF} $.

\subsection{Energy loss analysis}
\label{sec:Eloss_ana}

\begin{figure}[pos=h]
	\centering
	\includegraphics[width=1.0\linewidth]{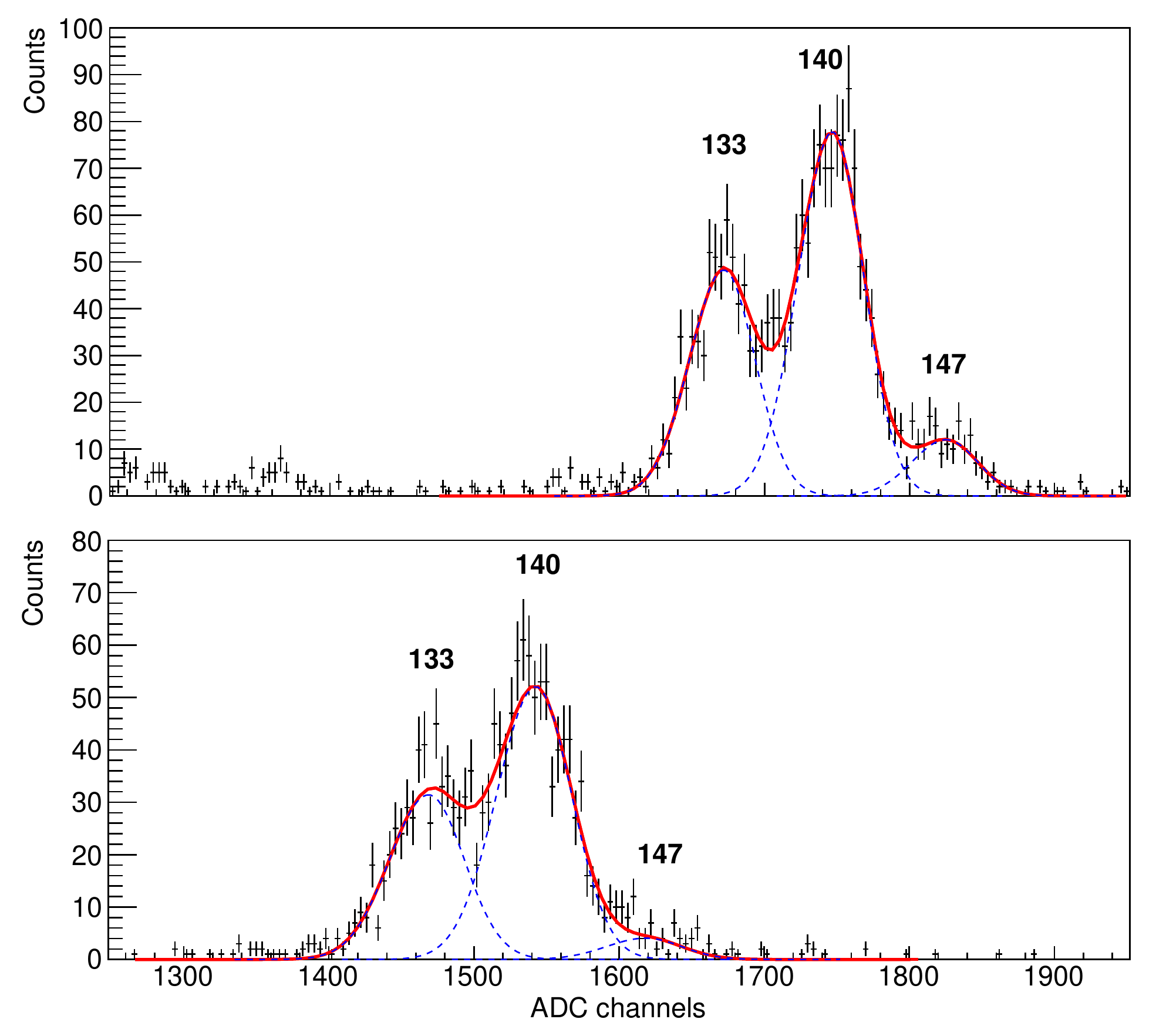}
	\caption{PIPS detector spectra for Lohengrin parameters set to A/q = 140/20 and E/q = 65/20. Top : no sample, Bottom: sample B in the beam. The acquisition time was identical for both measurements. }
	\label{fig:figexemple}
\end{figure}

The analysis of the spectra measured when the sample is placed in the beam is similar to the one for calibration data and, in particular, all peaks are fitted using asymmetric Gaussians. The main difference is that the process should account for energy loss in the sample to estimate the initial position of the peaks in the spectra. Figure \ref{fig:figexemple} illustrates that point on the energy loss of mass 140. The beam in that case is composed mainly of masses 133, 140 and 147. Stopping powers increase with mass, thus the 140 peak shifts more to low energy than the 133 one. This explains why peaks get closer when the sample is placed in the beam. We use SRIM tables to have a first estimate of the energy loss and of the peak position in the automatic fitting process. Since the area of the peaks does not change when the sample is placed in the beam, except for a common reduction explained by the fact that part of the beam is intercepted by the sample support, there is no doubt on peak attribution. 

\subsection{Stopping power determination}
\label{sec:Stop_power}

In the literature, the conversion from energy loss to stopping power is usually expressed by:
\begin{equation}
	\frac{-1}{\rho}\frac{dE}{dx}(E^*) =   \frac{E_0 - E_1}{\rho L}
	\label{eq:stop_pow_lit}
\end{equation}
with $E_0$, $E_1$ the energy of the beam before and after respectively crossing the sample thickness L, and $E^*$ sometimes set to $E_0$ or to $(E_0 + E_1) /2 $.

Here, and because our uncertainties were low enough to distinguish such differences, we used a more accurate conversion with the assumption that stopping power is linear over the small energy range between $E_0$ and $E_1$. In that case, the solution of the differential equation 
\begin{equation}
	\frac{dE}{dx}(E) = a\; E + b	
	\label{eq:stop_pow_2}
\end{equation}
is:
\begin{equation}
	E(x) = E_0 - \frac{(e^{-a x} -1) }{a} \frac{dE}{dx}(E_0) 
	\label{eq:stop_pow_3}
\end{equation}
then
\begin{equation}
	\frac{E_0 - E_1 }{L}=    \frac{e^{-a L}-1}{a\;L} \; \frac{dE}{dx}(E_0) 
	\label{eq:stop_pow_4}
\end{equation}
In the analysis process, we estimate parameter $a$ by fitting the slope of the variation of the experimental $\Delta E / L$ with energy. 
Finally, the mass stopping power is, using our previous notations ($E_0=E_L$, $E_1=E_R$ and $E_0 - E_1 = \Delta E$) :
\begin{equation}
	\frac{-1}{\rho}\frac{dE}{dL}(E_L) = \frac{aL}{1-e^{-a L}}\;\frac{\Delta E}{\rho L} 
\end{equation}

The thickness correction factor $ \frac{1-e^{-a L}}{aL} $ is not negligible in our data since it was found to range between 0.96 and 1 depending on the mass and sample thickness. 

In the same way as for the sample thickness determination in \ref{sec:sample_thickness}, an uncertainty of 0.4\% was added quadratically to the thickness uncertainty quoted in Table \ref{table:samplenm} to account for a possible misalignment of the foil in the beam by $\pm$5\degree.

\begin{figure}[pos=h]
	\centering
	\includegraphics[width=0.95\linewidth]{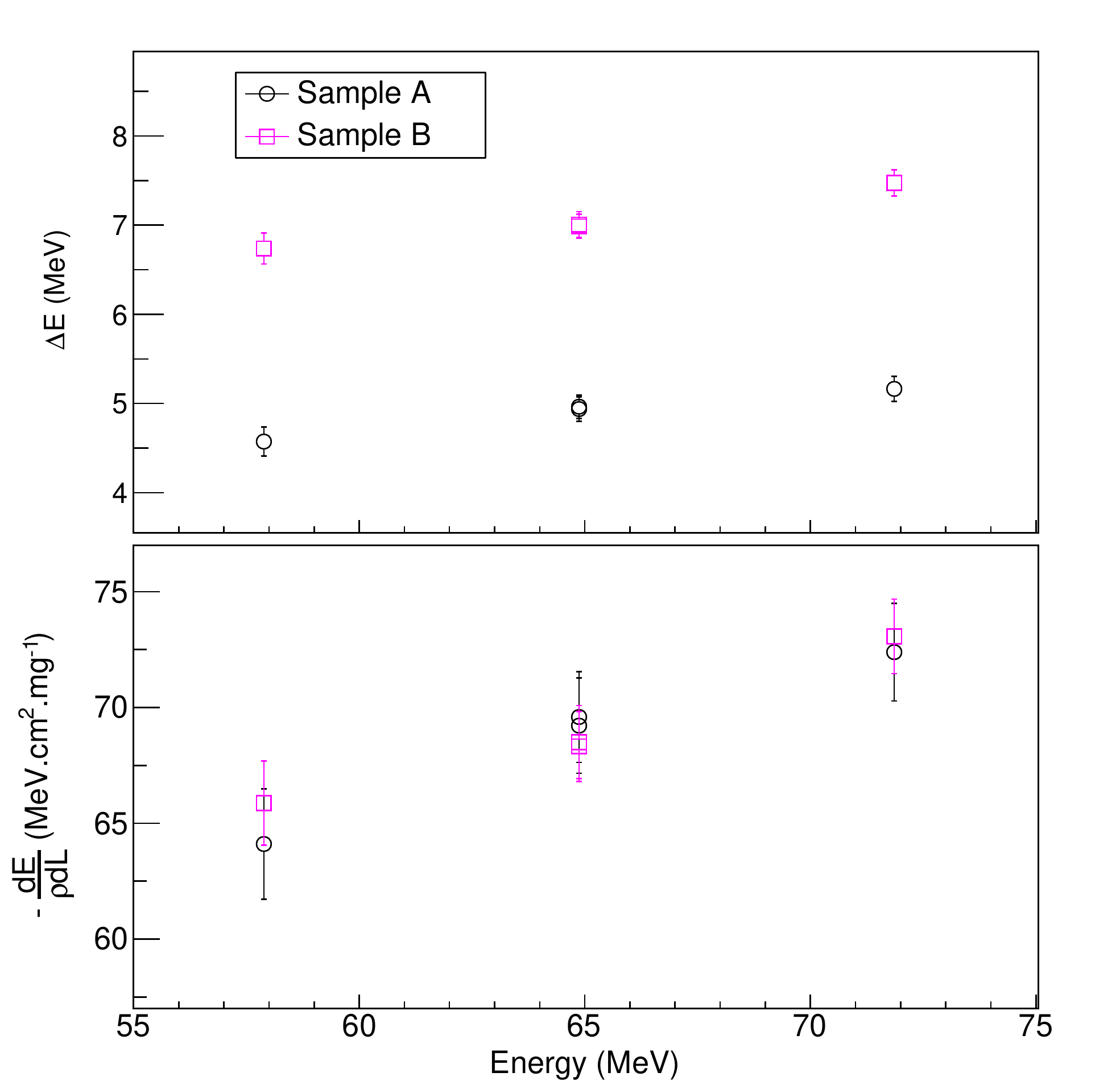}
	\caption{Top: energy losses obtained for fragments with mass 140 in the two Mylar samples as a function of energy. Bottom:corresponding calculated stopping powers.}
	\label{fig:140Cs_3mylars}
\end{figure}

Figure 	\ref{fig:140Cs_3mylars} shows the energy loss and the resulting stopping power for mass 140 in the two Mylar samples, A and B.  In that case, mass 140 was mainly composed of $^{140}$Xe and $^{140}$Ba (respectively 65\% and 35\%).  The correction factor was estimated from the slope of energy loss data to 0.981(7) in sample A and 0.972(7) in sample B. Stopping powers obtained from the two different samples match very well within their uncertainties.

\section{Results and Discussion}
\label{sec:6}

\subsection{Dependency with ionic charge}
It is known that the energy loss of fragments depends on their charge state. However this charge state is supposed to reach an equilibrium value rapidly when fragments cross matter \cite{Betz}. An ETACHA \cite{Lamour} calculation in LISE++ \cite{bazin_program_2002,tarasov_lise_2008} with a carbon target (Mylar not available in this code) and a bromium projectile of 105 MeV gave an equilibrium thickness of 30 $\mu$g/cm$^2$. The samples used in the present work have thicknesses much larger. The assumption of an equilibrated charge value seems to be valid. It is also important to verify that the energy loss does not depend of the charge state in our data. Indeed, Lohengrin selects a specific ratio of A/q. Therefore, if for a given A the energy loss varies with selected q, it is no more possible to compare directly the energy loss for different masses. 

Figure 12 shows the evolution of the energy loss as function of the selected charge state for A=100 at 100 and 90 MeV and for A=136 at 70 MeV. The weighted averages are also shown. It is clear that there is no strong dependence and thus, the charge state has not to be taken into account in the analysis.

\begin{figure}[pos=h]
	\centering
	\includegraphics[width=1.0\linewidth]{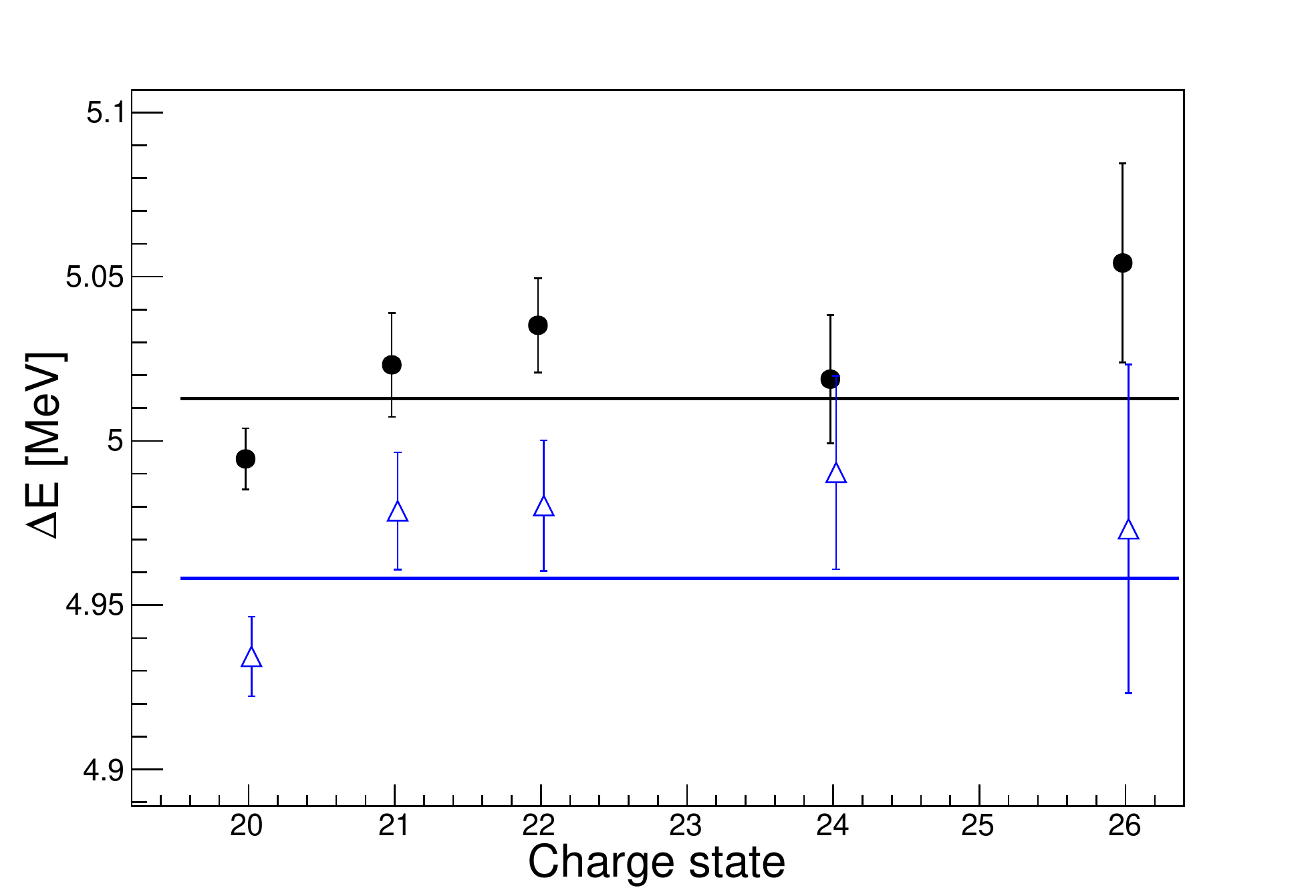}
	\includegraphics[width=1.0\linewidth]{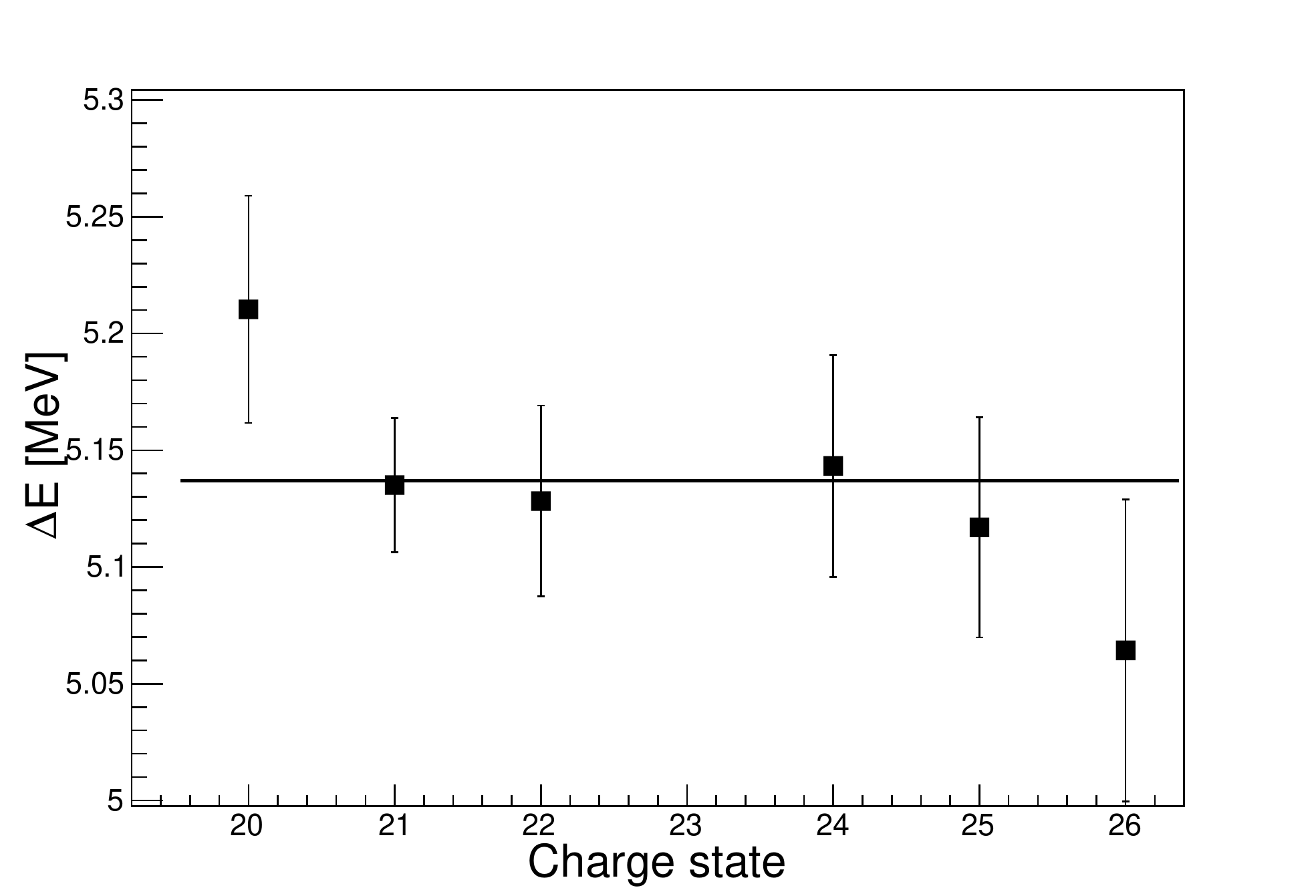}
	\caption{Energy losses as a function of the charge state. Top: A=100 for E=100 MeV (black points) and E=90 MeV (blue triangles). Bottom: A=136 and E=70 MeV. Lines on both figures correspond to the weighted average values. }
	\label{fig:}
\end{figure}

\subsection{Impact of the beam isobaric composition}
In order to compare our experimental results with the theoretical values obtained from models (Tables 8 and 9 in Appendix B), a good estimation of the isobaric composition or, alternatively, of the average nuclear charge of the fragment beam is mandatory.  

As explained in subsection \ref{sec:3.1}, the isobaric composition of fragments with mass A at the exit the Lohengrin spectrometer is directly related to their production by the fission process in the $^{239}$Pu target. Except for mass 140 and mass 144, the compositions were calculated from the JEFF-3.3 \cite{Plompen2020} fission yields for the $^{239}Pu(n_{th},f)$ reaction using :
\begin{equation}
	\label{eq.conc}
	C_A(Z) = Y(A,Z) \; / \sum_{Z'}{Y(A,Z')}
\end{equation}
where the sum runs over the few possible isobars with mass A, $C_A(Z)$ is the contribution of isobar Z in mass A and $Y(A,Z)$ is the parent independent fission yield for thermal neutrons of the fragment of mass A and atomic number Z. The average nuclear charge of the fragments with mass A, 	$\langle Z_A \rangle$ is obtained from : 
\begin{equation}
	\label{eq.Zav}
	\langle Z_A \rangle = \sum_{Z'}{Z' C_A(Z')}
\end{equation}

For mass 140 and 144, the isobaric composition is, as explained before, strongly perturbed by the presence of a ns-isomeric state. We observed in previous gamma-ray spectrometry measurements performed at Lohengrin that at high ionic charge setting beams of these two masses are essentially pure Cs \cite{Amouroux2013}.

One source of uncertainties on the isobaric composition is associated to the large uncertainties on the fission yields in JEFF-3.3 database \cite{Plompen2020}. We have estimated their impact on the average nuclear charge and stopping power calculations for two representative cases in Mylar:  with fragments of masses A = 100 and kinetic energies E = 100 MeV and with fragments of masses A = 139 and kinetic energies E = 69 MeV.

The fragments of mass 100 produced by the \break $^{239}Pu(n_{th},f)$ process are mainly $^{100}$Y, $^{100}$Zr , $^{100}$Nb and $^{100}$Mo with the following fission yields: 0.49 $\pm$ 0.12 $\%$, 4.22 $\pm$ 0.52 $\%$, 1.50 $\pm$ 0.36 $\%$ and 0.28 $\pm$ 0.10 $\%$. Their theoretical stopping power at 100 MeV in Mylar, according to the phenomenological model developed by Knyazheva et al., ranges from 65.75 to 70.36 MeV.cm$^2$/mg. The average nuclear charge for this isobaric composition, $ \langle Z_{100} \rangle $,  is about  40.24 and the theoretical stopping power for that composition is found to be about 67.66 MeV.cm$^2$/mg using:

\begin{equation}
	\frac{-1}{\rho }\frac{dE}{dL} (A) = \sum_Z{C_A(Z)*  \frac{-1}{\rho }\frac{dE}{dL} (Z,A) }
\end{equation}

Uncertainties on $ \langle Z_{100} \rangle $ and on the theoretical stopping power value were estimated to be 0.06 and 0.09 MeV.cm$^2$/mg, respectively, by using a Monte Carlo method that propagates fission yields uncertainties. The uncertainty on the theoretical stopping power is far much lower than our uncertainty on the experimental value, which is 0.8 MeV.cm$^2$/mg.

The same calculation was performed for fragments with masses 139. The average nuclear charge $ \langle Z_{139} \rangle $ is estimated to be 54.48 and its uncertainty to be 0.07. Their theoretical stopping power at 65 MeV is estimated to 67.34 MeV.cm$^2$/mg in Mylar (according to Knyazheva \emph{et al.} model) and its uncertainty to about 0.05 MeV.cm$^2$/mg. This later value is again much lower than the experimental uncertainty (here about 1.4 MeV.cm$^2$/mg).

We then concluded that fission yield uncertainties have a limited impact on the average nuclear charge ($\delta\langle Z_A \rangle/\langle Z_A \rangle \approx 0.1 \% $) and thus on the theoretical value of stopping powers. Therefore, we neglected them in the comparisons between data and models.

Another source of uncertainties regarding the isobaric composition concerns a possible bias due to the energy dependence. In our comparison to models, we made the hypothesis that the isobaric composition (and the average nuclear charge) of the fragments does not vary with their kinetic energy. This hypothesis is justified by the use in our measurement of a thick and quite old target, with a broad energy distribution that results from the superposition of fragments emitted from the different parts and depths of the target. This process should  wash out any differences in the energy distributions of individual isotopes.

However, since there is no way to estimate properly this process, we used the results of Schmitt et al. \cite{SCHMITT1984} (who measured the variation of the isobaric composition of light fragments with their energy at Lohengrin with a very thin target of $^{239}Pu$) to estimate the bias on the calculated stopping powers of light fragments. For mass 100, if one extrapolates that the nuclear charge varies between 40.5 (at 70 MeV) and 40.0 (at 107 MeV) and compares it to the case where the nuclear charge is fixed at 40.21over the same range, the bias on the stopping power calculated according to Knyazheva \emph{et al.} ranges between -0.3 and 0.3 MeV.cm$^2$.mg$^{-1}$. Such bias is about 3 times smaller than the experimental uncertainties of our measured stopping powers.

The same analysis can not be performed with heavy fragments because no precise measurement of the variation of the average nuclear charge with the kinetic energy exists. On the other hand, a good estimation of this variation can be obtained from simulations with the GEF fission code \cite{Schmidt2016}. This code was tested to provide consistent nuclear charge variations with Schmitt ones \cite{SCHMITT1984} for fragments with masses 95, 100 and 105. GEF simulations show that the average nuclear charge of the fragments of masses 139 (mainly composed of I, Xe, Cs and Ba) decreases rather linearly from 55 to 54 over the large part of the kinetic energy distribution.  Using a fixed nuclear charge at 54.5 over the same range introduces a bias in the stopping power that varies between -0.4 and 0.4 MeV.cm$^2$.mg$^{-1}$. Such bias is in most cases 3 times smaller than the experimental uncertainties on the stopping powers of heavy fragments.

We therefore neglected any dependency of the isobaric composition with the kinetic energy of the fragments in our comparisons between data and models.

\begin{figure}[pos=h]
	\includegraphics[width=0.98\columnwidth]{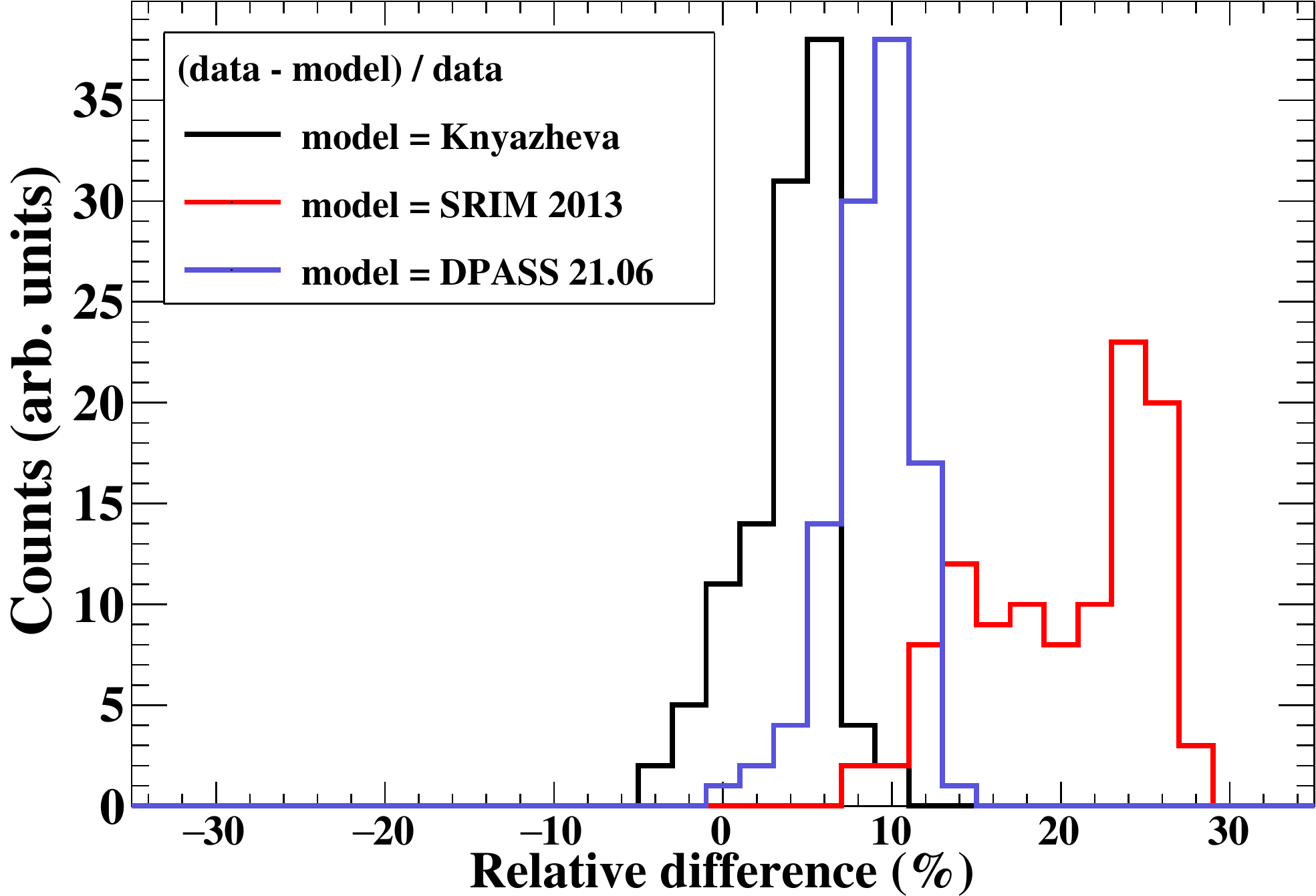} \\
	\includegraphics[width=0.98\columnwidth]{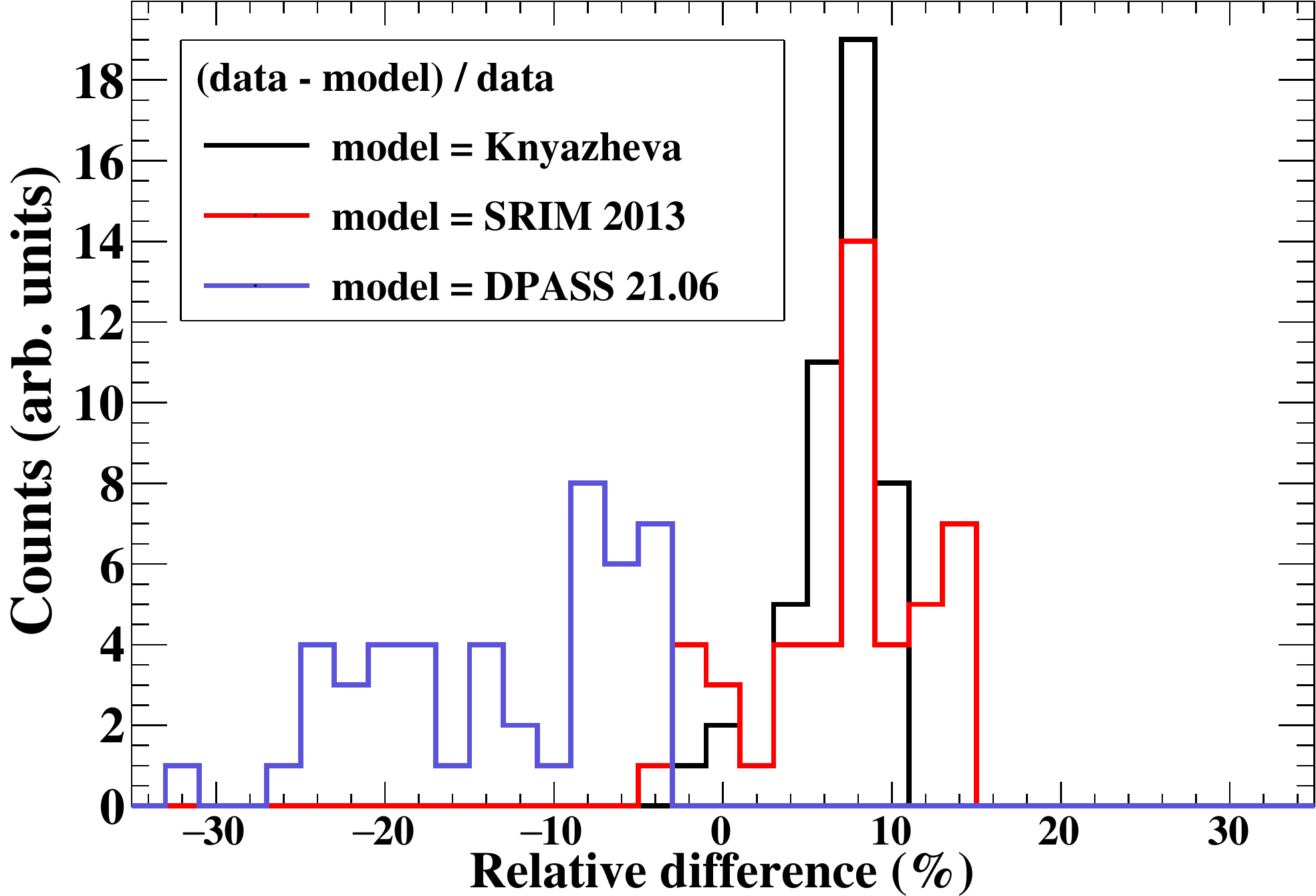}	
	\caption{\underline{Top:} relative difference distribution between our data on Mylar sample and Knyazheva \emph{et al.} \cite{KNYAZHEVA2006} model (black), DPASS 21.06 \cite{dpass} database (blue) and SRIM 2013 \cite{ziegler_srim_2010} (red).
	\underline{Bottom:} same as top for nickel sample.}
	\label{fig:Rel-Diff}
\end{figure}

\subsection{Stopping power comparison with models}

Stopping powers were obtained for 24 masses in Mylar and 12 masses in nickel. They are detailed in Tables \ref{tab:results_mylar} and \ref{tab:results_Ni} in Appendix \ref{app1}. Relative uncertainties on stopping powers range between 0.6\% and 4\% and are most often dominated by the detector instabilities detailed in subsection \ref{sec:Detec_stab}. 

Figure \ref{fig:Rel-Diff} shows an overall comparison between our data and 3 different models:
\begin{itemize}
    \item the Knyazheva \emph{et al.} \cite{KNYAZHEVA2006} model, chosen because it has been especially tuned for fission fragment energy loss.
    \item the SRIM-2013 \cite{ziegler_srim_2010} code, chosen because it is widely used.
    \item the DPASS 21.06 \cite{dpass} database, chosen because it is at the state of the art for describing  heavy ions energy loss in the Bragg peak region.
\end{itemize}

For the Mylar sample (Fig. \ref{fig:Rel-Diff} top), our data are in average 5\% higher than Knyazheva \emph{et al.} \cite{KNYAZHEVA2006}, and in perfect agreement if one select the light group fragment only (see Fig. \ref{fig:MylarKnya}). The shape of the relative difference distribution with DPASS database is very similar to the shape with Knyazheva but shows an average discrepancy of about 10\%. On the other hand the comparison with SRIM-2013 \cite{ziegler_srim_2010} is widely distributed from 10 to 30\% without any clear trend.

For the nickel sample (Fig.\ref{fig:Rel-Diff} bottom), our data are in average 7 to 8\% higher than Knyazheva \emph{et al.} \cite{KNYAZHEVA2006}. A comparison with SRIM-2013 \cite{ziegler_srim_2010} gives a similar average value but a slightly broader distribution. The comparison to DPASS 21.06 database \cite{dpass} is widely distributed from 0 to -30\%.

\subsection{Models adjustment to our data}

In order to be able to use these data for our fission applications, where one need to cover the full range in energy and mass associated with fission, a fit based on 2 previously described models have been performed. The fit is bidimensional, \emph{i.e.} the energy and mass (Z) dependence are considered simultaneously. The Z from our data has been determined by taking the composition for each mass given by the JEFF-3.3 \cite{Plompen2020} fission yields as explained above, nevertheless we have checked that taking an average Z instead has a negligible, \emph{i.e.} well below the percent level, impact on the calculated value. 

\begin{figure}[pos=h]
	\includegraphics[width=0.98\columnwidth]{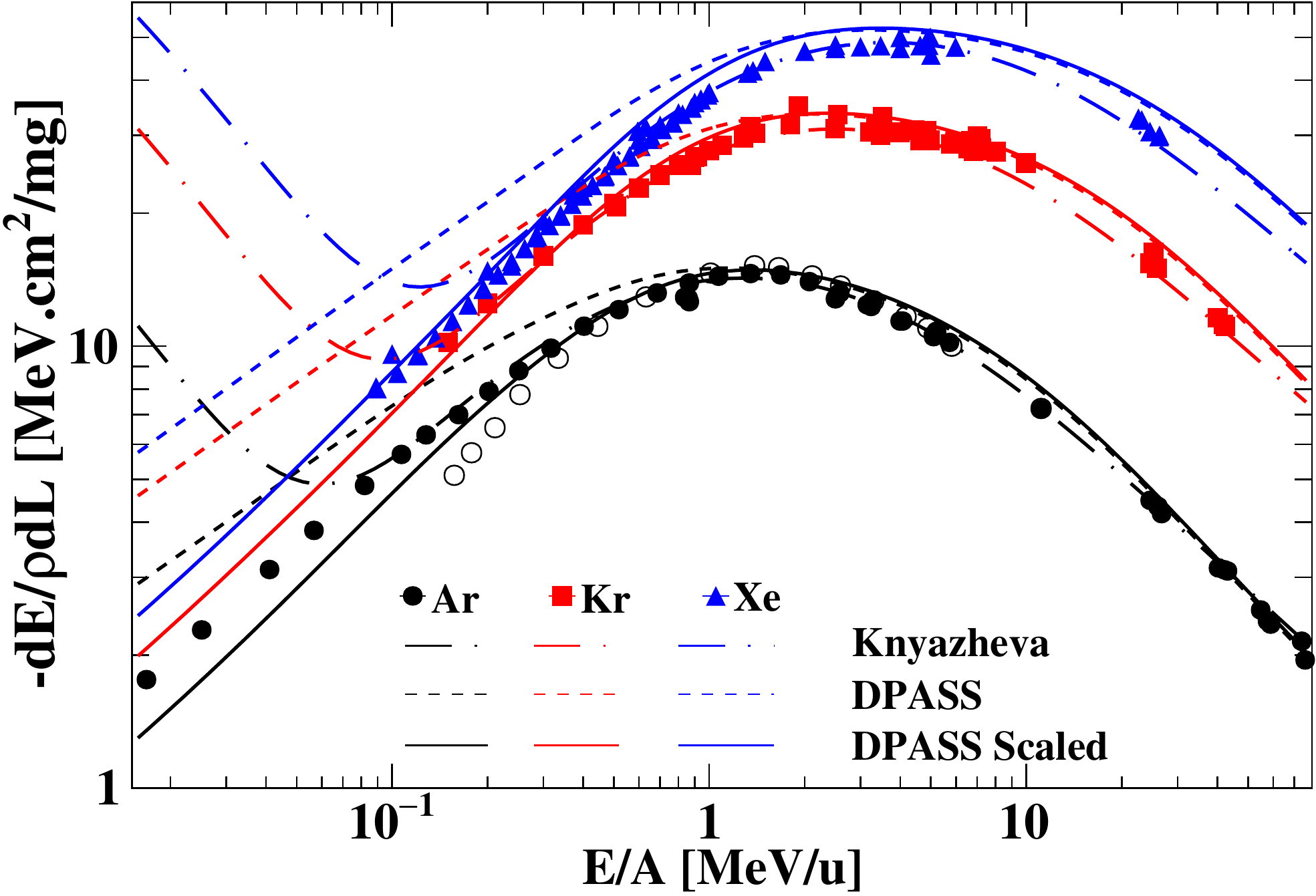}
	\caption{Ar, Kr and Xe stopping powers on nickel data \cite{iaea-nds-stopping} (symbols) compared to Knyazheva \emph{et al.} \cite{KNYAZHEVA2006} model (dot-dashed lines) and to DPASS \cite{dpass} database before (dash lines) and after scaling to our data (solid lines)}
	\label{fig:IAEA-DPASS-Ni}
\end{figure}

The first model is based on the DPASS 21.06 library \cite{dpass}, briefly described in section \ref{sec:2}. To perform a fit starting from these tabulated values, the stopping power is calculated by using linear interpolation from (log(E/A),~Z$_1$,~$\frac{- \, dE}{\rho \, dL \; Z_1}$) tables. 

Figure \ref{fig:IAEA-DPASS-Ni} shows a comparison of the DPASS 21.06 database (dash line) with the stopping power of Ar, Kr and Xe ions on nickel extracted from the IAEA database of experimental data \cite{iaea-nds-stopping}. Below the Bragg peak, one can notice a significant deviation of the DPASS database from the experimental data. Since in this energy region the stopping power has a nearly exponential dependence in E/A, we arbitrarily chose to scale the DPASS database using the following equation: 
\begin{eqnarray}
	\left. \frac{dE}{\rho dL}\right|_{Scaled}  =  \frac{N}{1 + S \exp{\frac{-v}{v_0 Z_1^{1/6}}}} \, 
	\left. \frac{dE}{\rho dL}\right|_{DPASS}
	\label{eq:dpass-scaling}
\end{eqnarray}
where $N$ is an overall normalisation parameter and the parameter $S$ allows to modify the energy dependence below the Bragg peak region. The solid line in Fig. \ref{fig:IAEA-DPASS-Ni} shows the scaled stopping power using Eq. \ref{eq:dpass-scaling} fitted to reproduce our data on nickel shown hereafter. 

The second chosen model is the one developed by Knyazheva \emph{et al.}  \cite{KNYAZHEVA2006} briefly described in section \ref{sec:2}. This model has 4 adjustable parameters per target: $\lambda$ associated with the effective nuclear charge of the projectiles, $\overline{Z_1}$, and 3 parameters $a_0$ to $a_2$ associated with the $L_{Bohr}$ power series expansion ($L_{Bohr}^0$ to $L_{Bohr}^2$). The initial parameters for Mylar and nickel foils are taken from \cite{KNYAZHEVA2006}. Dot-dashed lines in Fig. \ref{fig:IAEA-DPASS-Ni} show a comparison of Knyazheva \emph{et al.} model results with the stopping power of Ar, Kr and Xe ions on nickel extracted from the IAEA database of experimental data \cite{iaea-nds-stopping}. The agreement with data is excellent in the energy region of interest for fission fragments, but at variance below about 0.2 MeV/u (\emph{i.e.} below the fission fragment ROI in energy).

\begin{figure*}
	\centering
	\includegraphics[width=0.85\textwidth]{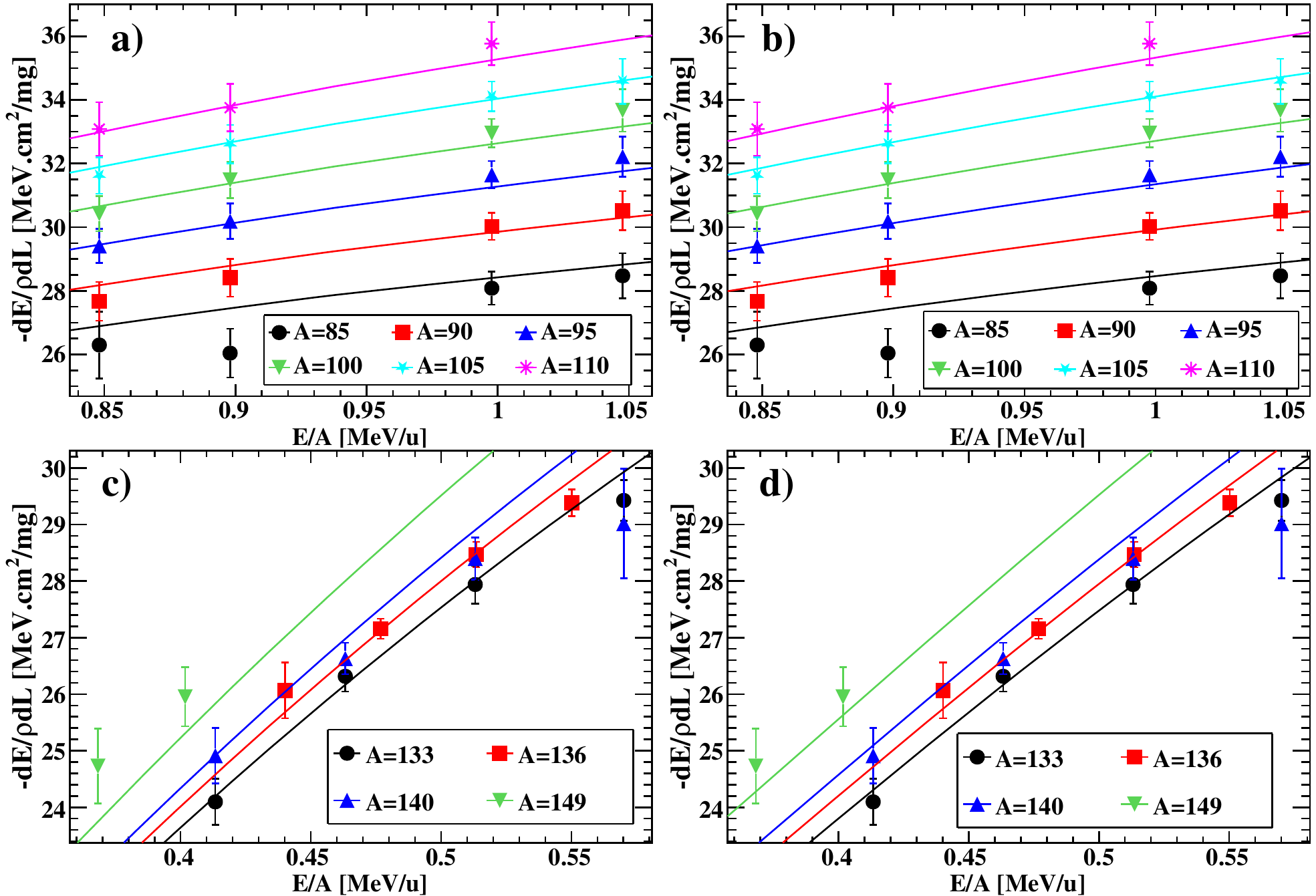}
	\caption{Stopping powers obtained in this work with a nickel sample (symbols) for light (a,b) and heavy fragments (c,d). The scaled DPASS (eq. \ref{eq:dpass-scaling}) (a,c) and the fitted Knyazheva \emph{et al.} \cite{KNYAZHEVA2006} (b,d) models are also plotted (solid lines).}
	\label{fig:NiRes}
	\centering
	\vspace{0.8cm}
	\includegraphics[width=0.85\textwidth]{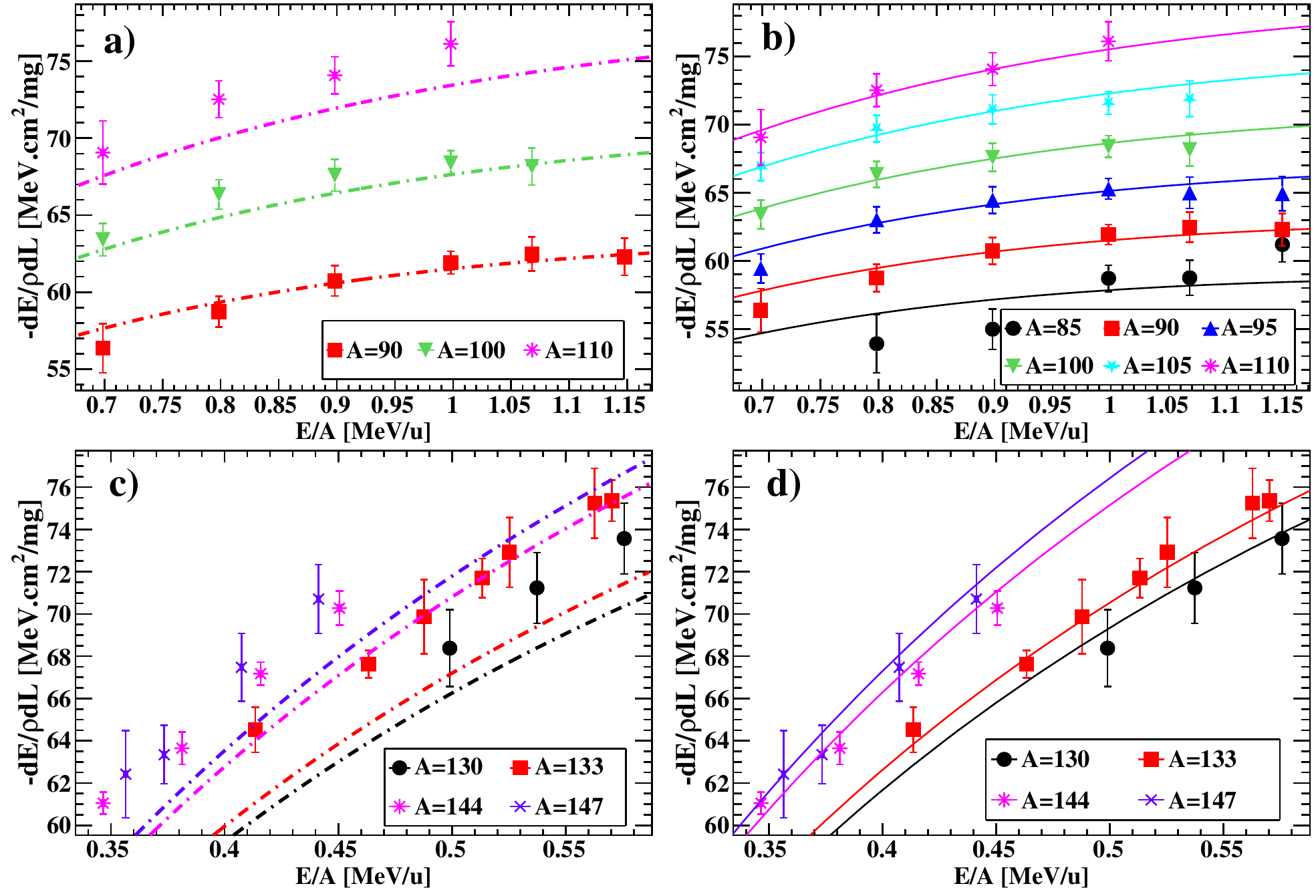}
	\caption{Stopping powers obtained in this work on Mylar samples (symbols) for light (a,b) and heavy fragments (c,d) and compared to the Knyazheva \emph{et al.} \cite{KNYAZHEVA2006} model. The dash-dotted line in (a,c) corresponds to the initial model values. Solid lines in (b,d) are the best fit of the data.}
	\label{fig:MylarKnya}
\end{figure*}

Data for the nickel sample are shown in Fig. \ref{fig:NiRes}. Data for light mass fragments (subset a and b) show a smooth increase with mass and energy. Data for heavy mass fragments (subset c and d) follow a similar trend. Nevertheless the stopping power mass dependence is much weaker for the heavy group, and does not seem as smooth as for the light mass fragments group.

The best fit results for our nickel data using the Knyazheva model is reported in Table \ref{tab:BestFitKnya} and shown in the right panels (b \& d) of Fig. \ref{fig:NiRes}. The reduced $\chi^2$ corresponding to this fit is $\chi^2/\mbox{NDF} \, = \, 38.5/39$ denoting the excellent agreement between our data and the fit results.

\begin{table}[pos=h]
	\caption{Best fit parameters of our data using Knyazheva \emph{et al.} \cite{KNYAZHEVA2006} model.}
	\begin{tabular}{|c|rcl|rcl|}
		\hline
		& \multicolumn{3}{|c|}{nickel} & \multicolumn{3}{|c|}{Mylar} \\ \hline
		$\lambda$ &     0.520 &$\pm$& 0.030  & 0.489 &$\pm$& 0.025\\ 
		$a_0$    &   0.97 &$\pm$ & 0.12 &  0.73  &$\pm$& 0.10\\ 
		$a_1$     &  -0.535 &$\pm$& 0.032 &  -0.605 &$\pm$& 0.040\\ 
		$a_2$      &  0.069 &$\pm$& 0.0044  & 0.049 &$\pm$& 0.010 \\ \hline
	\end{tabular}
	\label{tab:BestFitKnya}
\end{table}

One may notice a strong correlation {(see Table \ref{tab:CorrKnyaNickel}} between most of the parameters, probably due to our limited region of interest to constraint the fit.

The best fit result of our nickel data based on Eq. \ref{eq:dpass-scaling} is given on Table \ref{tab:BestFitDPASS} and shown in the left panels (a \& c) of Fig. \ref{fig:NiRes}.
The reduced $\chi^2$ corresponding to this fit is $\chi^2/\mbox{NDF} \, = \, 41.7/41$ denoting the excellent agreement between our data and the scaling adjustment.

\begin{table}[pos=h]
	\centering
	\caption{Best fit parameters of our data using a scaling (eq. \ref{eq:dpass-scaling}) of DPASS 21.06 library \cite{dpass}.}
	\begin{tabular}{|c|rcl|rcl|}
		\hline
		& \multicolumn{3}{|c|}{nickel} & \multicolumn{3}{|c|}{Mylar} \\ \hline
		$N$ &     1.0139 &$\pm$& 0.0066  & 1.0992 &$\pm$& 0.0017\\ \hline
		$S$    &   2.09 &$\pm$ & 0.10 & \multicolumn{3}{|c|}{0 (fixed)}\\ \hline
	\end{tabular}
	\label{tab:BestFitDPASS}
\end{table}

The result of this fit is also shown as solid line in Fig \ref{fig:IAEA-DPASS-Ni}. One can notice that the scaling term based on our data also allows to retrieve a good trend for the scaled DPASS library with respect to experimental data available in the IAEA data\-base \cite{iaea-nds-stopping} for nickel. 

Data for Mylar samples (Table \ref{tab:results_mylar}) are shown and compared to Knyazheva \emph{et al.} \cite{KNYAZHEVA2006} fitted model in Fig. \ref{fig:MylarKnya} and using Eq. \ref{eq:dpass-scaling} of Fig. \ref{fig:MylarDPASS}. Since Mylar is a compound material (C$_{10}$H$_{8}$O$_4$) Bragg sum rule was used in model calculations, assuming negligible compound corrections \cite{sigmund_valence_2005} for incoming fission fragments.
The overall trend of the data is very similar to the one on nickel discussed above.

Figure  \ref{fig:MylarKnya} shows a comparison with Knyazheva \emph{et al.} \cite{KNYAZHEVA2006} before (dash-dotted lines) and after fitting (solid lines). One has a perfect agreement with Knyazheva \emph{et al.} \cite{KNYAZHEVA2006} data for the lightest mass, and a growing deviation with increasing mass to reach about 5\% for the heavier masses. After fitting  Fig. \ref{fig:MylarKnya}(b,d) shows a perfect agreement between our data and  Knyazheva \emph{et al.} \cite{KNYAZHEVA2006} model.
The best fit result of our Mylar data using   Knyazheva \emph{et al.} \cite{KNYAZHEVA2006} model is  given in table \ref{tab:BestFitKnya}.
The reduced $\chi^2$ corresponding to this fit is $\chi^2/\mbox{NDF} \, = \, 55.7 / 94$  denoting the excellent agreement between data and fitted results. 

As in the nickel case, because of our limited range in energy, one has a strong correlation (see table \ref{tab:CorrKnyaMylar}) between all the fitting parameters. 

\begin{figure}[pos=h]
	\includegraphics[width=0.98\columnwidth]{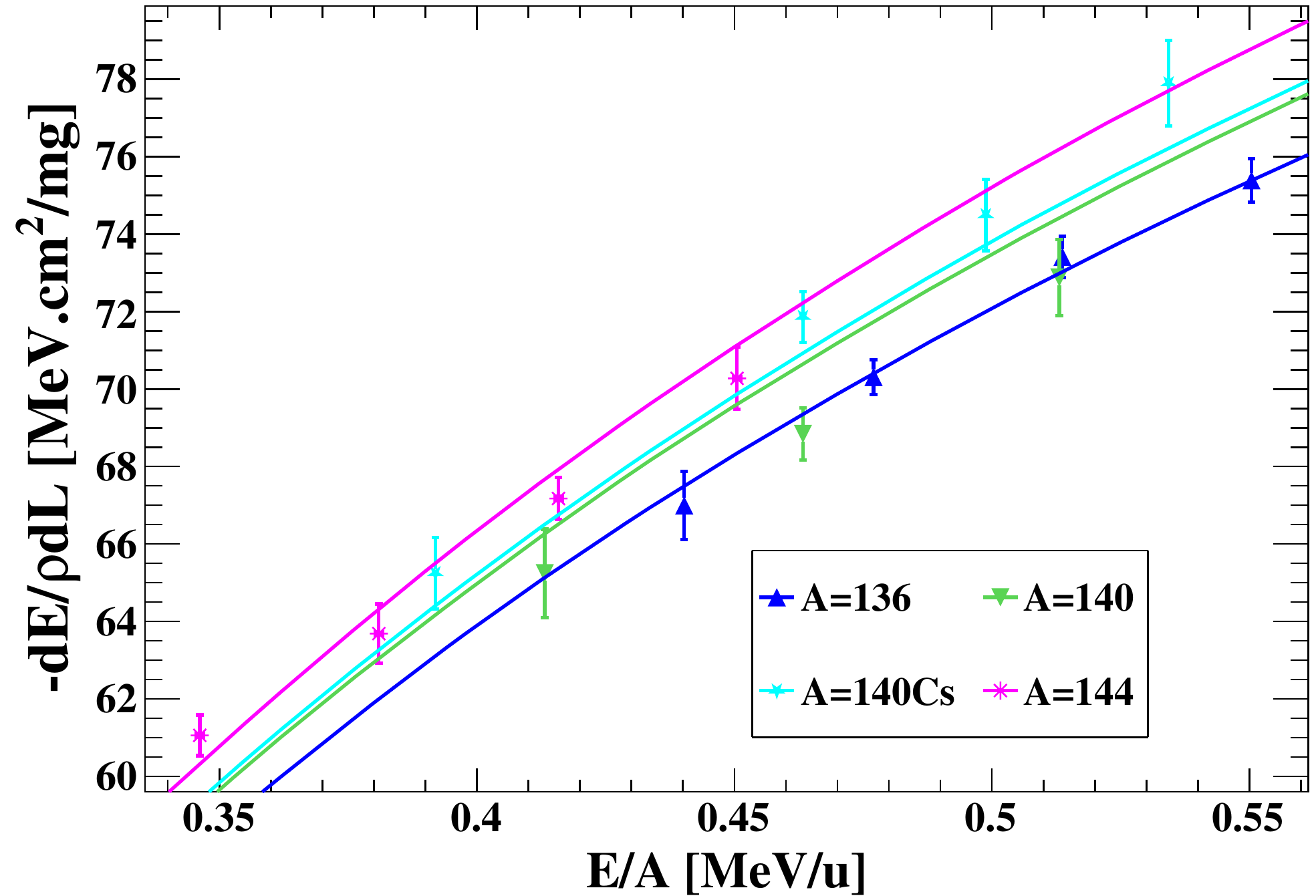}
	\caption{Stopping powers obtained in this work on Mylar samples (symbols) for $^{140}$Cs and surrounding masses compared to our fit using Knyazheva \emph{et al.} \cite{KNYAZHEVA2006} model (solid lines and corresponding colors).}
	\label{fig:CsMylarKnya}
\end{figure}

Figure \ref{fig:CsMylarKnya} shows the comparison between our data on Mylar and our best fit using Knyazheva \emph{et al.} \cite{KNYAZHEVA2006} model around the mass 140. In this very limited mass region, one may notice 
that the model well reproduces masses 136 and 144 but a clear discrepancy appears for mass 140. Whereas $^{140}Cs$ is slightly higher but in the 1 sigma uncertainty band, mass 140 is too low. 
It is not clear if this small discrepancy comes from the data (uncorrected remaining systematic bias, wrong assumption in the Z distribution associated with mass 140 
or if the model needs some small refinements in the iodine to baryum region.

\begin{figure*}
	\centering
	\includegraphics[width=0.85\textwidth]{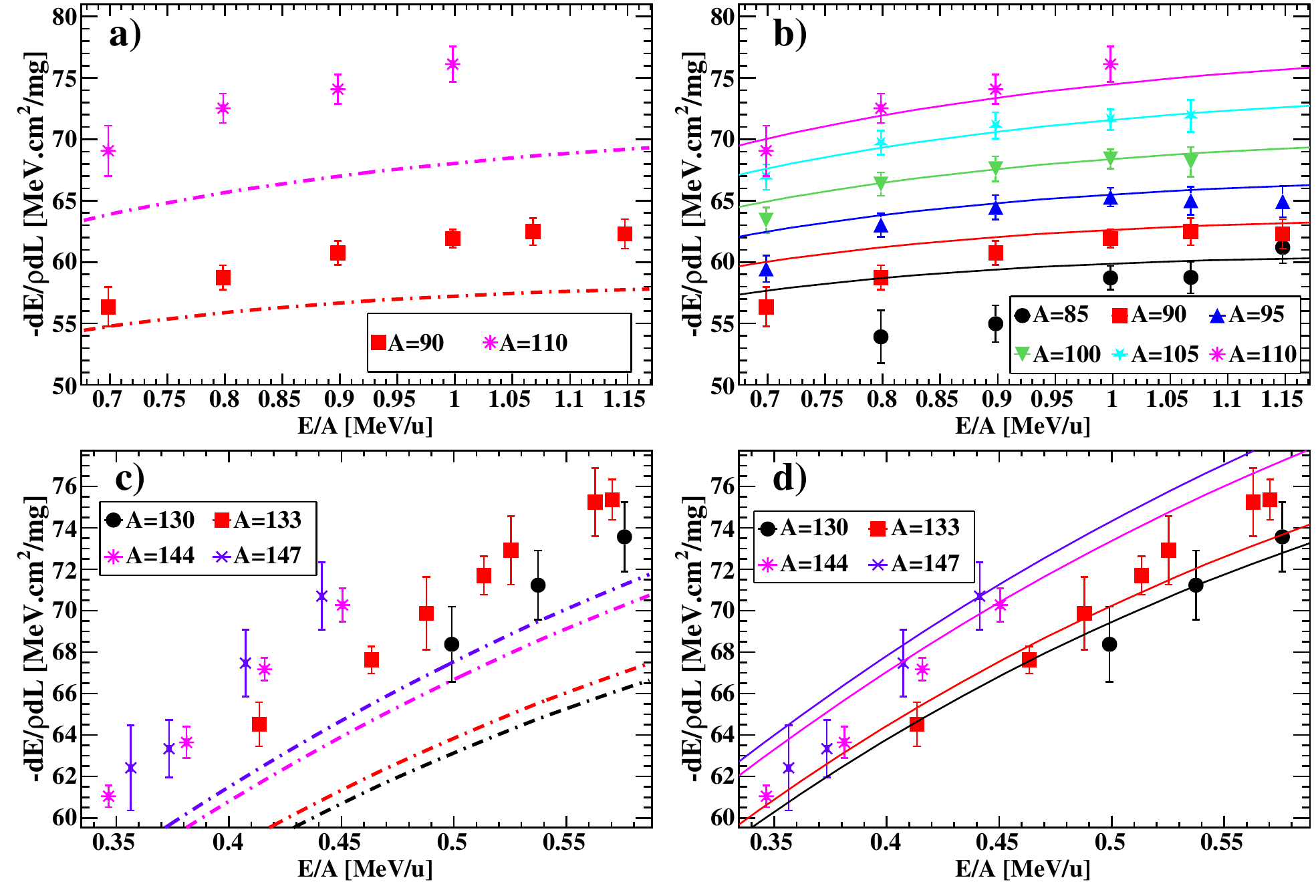}
	\caption{Stopping powers obtained in this work on Mylar samples (symbols) for light (a,b) and heavy fragments (c,d) fitted using Eq. \ref{eq:dpass-scaling}. The dash-dotted line in (a,c) corresponds to the initial model values. Solid lines in (b,d) are the best fit of the data.}
	\label{fig:MylarDPASS}
\end{figure*}

Figure \ref{fig:MylarDPASS} subset (a,c) shows the comparison between our data on Mylar and the DPASS database \cite{dpass} (dash-dotted lines) and subset (b,d) after scaling using Eq. \ref{eq:dpass-scaling}. 
Our whole Mylar data are systematically about 10\% higher than the \allowbreak DPASS database \cite{dpass}. So, in the fit using Eq. \ref{eq:dpass-scaling} the parameter $S$ was fixed to 0\, since no energy dependence of scaling seems necessary in this case. The best fit results shown in Fig. \ref{fig:MylarDPASS} are reported in Table \ref{tab:BestFitDPASS}.

The reduced $\chi^2$ is $\chi^2/\mbox{NDF}  =  165.0 / 97$ indicating a satisfactory, but not perfect, agreement between our Mylar data and the scaled DPASS database. In particular, one can notice in Fig. \ref{fig:MylarDPASS}(b,d) that the slope in energy is systematically slightly higher in our data in comparison with the scaled DPASS database.

To summarize, our data are in overall good agreement (\emph{i.e.} better than 10\%) for both nickel and Mylar with the Knyazheva \emph{et al.} \cite{KNYAZHEVA2006} model. A small discrepancy is still observed with increasing mass of the fragments. This could either be due to a different Z assumption as a function of mass, or to a worse mass resolution in case of the Knyazheva \emph{et al.} \cite{KNYAZHEVA2006} data.
After fitting, the agreement with our whole data set is excellent. Unfortunately, as shown in Fig. \ref{fig:IAEA-DPASS-Ni}, this model is only valid above 0.2~MeV/u that might be problematic in case of significant energy loss of fragments in few $\mu$m foils.

\section{Conclusions} 
\label{sec:7}

Accurate energy losses have been measured for a large range of fission fragments and incident kinetic energies in thin Mylar and nickel foils using the Lohengrin mass separator at ILL. The samples have been characterized precisely in thickness thanks to the alpha transmission method. 

Many settings (A/q, E/q) of the Lohengrin spectrometer were used, allowing the study for both light (6 different masses) and heavy (18 different masses for Mylar, 6 different masses for nickel) fission fragments. These ions were studied over a large range in incident energies: 60-110 MeV and 55-75 MeV for light and heavy fragments respectively. 

The experimental procedure and the analysis methodology have been extensively described. A special attention has been paid to the uncertainty calculations and correlation matrices. 

Energy losses were used to calculate the stopping powers in nickel and Mylar. For light fragments, a smooth evolution of the stopping powers as a function of the incident velocity is observed. For a given velocity, the energy loss increases also smoothly with the mass. For heavy fragments, the behavior is quite different. As expected, the increase of the stopping power is larger for a given mass as a function of the velocity. But for a given velocity, the stopping power is almost the same for different heavy masses.

Our experimental stopping power results have been compared to three models: the phenomenological model of \break Knyazheva \emph{et al.}, the DPASS 21.06 semi-empirical approach and the SRIM-2013 \cite{ziegler_srim_2010} code. It was found an overall good agreement, within 10$\%$, for the phenomenological model of \break Knyazheva for both Mylar and nickel samples. To note that our data are systematically slightly higher than the model. DPASS 21.06 slightly underestimates (with an average of 10 \% shift) the stopping power compared to our data for Mylar and overestimate it for nickel with a larger dispersion. SRIM-2013 \cite{ziegler_srim_2010} code slightly underestimates (with an average of 10 \% shift) the stopping power for nickel and largely underestimates it for Mylar.

In the range of our study, it validates the use of the \allowbreak Knyazheva \emph{et al.} and DPASS 21.06 models for accurate energy loss calculations with fission fragments until a predictive model is proposed. In order to provide references, both models parameters were fitted on our data. 

The DPASS data library, with the appropriate rescaling for nickel and Mylar, was integrated in the stopping power libraries of GEANT4 for the purpose of the FALSTAFF collaboration. In addition to simulations, this library is useful for data analysis in order to obtain the energy at the fission point.

Finally, while this work was intended to produce more precise stopping power values required in the development of fission fragment spectrometers, we believe that our results, where we used well characterized exotic nuclei as projectile, may be as well applicable for benchmarking the theory on energy loss in matter.

\section{Acknowledgements}
The authors would like to specially thank the P2IO LabEx (ANR-10-LABX-0038) in the framework "Investissements d’Avenir" (ANR-11-IDEX-0003-01) managed by the Agence Nationale de la Recherche (ANR, France) for making this work possible.
We are grateful to Vincent Morel and Patrick Champion for their help to adapt the reaction chamber.
We thanks Peter Sigmund for providing useful informations related to DPASS.

\nolinenumbers

\pagebreak

\bibliographystyle{elsarticle-num}
\bibliography{allreferences}

\appendix

\section{Correlation matrices}

\begin{table}[pos=h]
	\centering
	\caption{Correlation matrix for our best fit on nickel data using the Knyazheva \emph{et al.} \cite{KNYAZHEVA2006} model.}
\begin{tabular*}{\columnwidth}{c| c c c c}
	\hline
             & $\lambda$ & $a_0$ & $a_1$ & $a_2$ \\
             \hline
$\lambda$ &     1  &   0.999  &   0.943   & 0.016 \\
$a_0$    &     0.999     &    1   &  0.956  &   0.053 \\
$a_1$     &   0.943    &  0.956    &      1   &  0.339 \\
$a_2$      &        0.016 &    0.053  &   0.339    &      1 \\
	\hline	
\end{tabular*}
\label{tab:CorrKnyaNickel}
\end{table}

\begin{table}[pos=h]
\caption{Correlation matrix for our best fit on nickel data using a scaling (eq. \ref{eq:dpass-scaling}) of the DPASS 21.06 library \cite{dpass} }
\begin{tabular*}{\columnwidth}{c|cc}
	\hline
           & 	$N$ &        $S$ \\
           	\hline
$N$ &    1    & 0.934 \\
$S$ &   0.934    &       1 \\
	\hline	
\end{tabular*}
\label{tab:CorrDPASSNickel}
\end{table}

\begin{table}[pos=h]
	\caption{Correlation matrix for our best fit on mylar data using the Knyazheva \emph{et al.} \cite{KNYAZHEVA2006} model.}
	\begin{tabular*}{\columnwidth}{c | c c c c}
	\hline	
    & $\lambda$ & $a_0$ & $a_1$ & $a_2$ \\
\hline
$\lambda$ &     1  &   1 &     0.991 &   -0.893\\
$a_0$    &  1 &       1  &   0.991   & -0.895 \\
$a_1$     &   0.991 &     0.991 &           1  &   -0.936  \\
$a_2$      &    -0.893 &    -0.895 &     -0.936 &           1 \\
	\hline	
\end{tabular*}
\label{tab:CorrKnyaMylar}
\end{table}

\clearpage

\onecolumn
\section{Stopping power tables for Mylar and nickel}
\label{app1}

\begin{longtable}{c c  c  c c c c c c}
	
	\caption{Stopping power in Mylar obtained in this work, compared to the values of SRIM-2013 \cite{ziegler_srim_2010}, DPASS 21.06 \cite{schinner_expanded_2019} and the Knyazheva et al. model \cite{KNYAZHEVA2006}. The last two columns shows the rescaled DPASS values and the fitted  Knyazheva et al.  model respectively. }
	\label{tab:results_mylar}\\
	\hline
	\multirow{2}{*}{Mass} & \multirow{2}{*}{<Z>} & \multirow{2}{*}{Energy (MeV)} & \multicolumn{6}{c}{Stopping power (MeV.cm$^2$.mg$^{-1}$)}  \\
	& & & This work & TRIM & DPASS & Knyazheva & Fit DPASS & Fit Knyazheva \\
	\hline
	\endfirsthead
	
	\multicolumn{9}{l}%
	{{\tablename\ \thetable{} -- continued from previous page}} \\\\
	\hline
	\multirow{2}{*}{Mass} & \multirow{2}{*}{<Z>} & \multirow{2}{*}{Energy (MeV)} & \multicolumn{6}{c}{Stopping power (MeV.cm$^2$.mg$^{-1}$)}  \\
	& & & This work & TRIM & DPASS & Knyazheva & Fit DPASS & Fit Knyazheva \\
	\hline
	\endhead
	\endfoot
	
	\hline \hline
	\endlastfoot

  & & 67.9 & 53.8 $\pm$ 2.1 & 49.3 & 53.6 & 56.4 & 58.8 & 56.3\\
 & & 76.3 & 54.9 $\pm$ 1.5 & 50.5 & 54.2 & 57.5 & 59.5 & 57.2\\
 & & 84.8 & 58.7 $\pm$ 1.0 & 51.4 & 54.7 & 58.3 & 60.0 & 57.9\\
 & & 90.8 & 58.7 $\pm$ 1.3 & 51.9 & 55.0 & 58.7 & 60.3 & 58.2\\
 \multirow{-5}{*}{85} & \multirow{-5}{*}{34.3} &97.6 & 61.1 $\pm$ 1.3 & 52.3 & 55.1 & 59.0 & 60.5 & 58.4\\
 \hline
 & & 62.9 & 56.3 $\pm$ 1.6 & 51.0 & 54.7 & 57.6 & 60.1 & 58.0\\
 & & 71.9 & 58.7 $\pm$ 1.0 & 52.4 & 55.9 & 59.3 & 61.3 & 59.6\\
 & & 80.8 & 60.7 $\pm$ 1.0 & 53.4 & 56.7 & 60.6 & 62.2 & 60.7\\
 & & 89.8 & 61.9 $\pm$ 0.7 & 54.1 & 57.2 & 61.5 & 62.8 & 61.5\\
 & & 96.1 & 62.4 $\pm$ 1.1 & 54.5 & 57.5 & 62.0 & 63.1 & 61.9\\
 \multirow{-6}{*}{90} & \multirow{-6}{*}{36.3} &103.3 & 62.2 $\pm$ 1.2 & 54.8 & 57.7 & 62.4 & 63.4 & 62.2\\
 \hline
 & & 66.4 & 59.4 $\pm$ 1.1 & 51.5 & 57.0 & 60.3 & 62.5 & 61.0\\
 & & 75.9 & 62.9 $\pm$ 0.9 & 53.3 & 58.2 & 62.1 & 63.9 & 62.8\\
 & & 85.3 & 64.4 $\pm$ 1.0 & 54.8 & 59.2 & 63.6 & 64.9 & 64.1\\
 & & 94.8 & 65.2 $\pm$ 0.8 & 55.9 & 59.8 & 64.6 & 65.7 & 65.1\\
 & & 101.5 & 64.9 $\pm$ 1.1 & 56.6 & 60.2 & 65.2 & 66.1 & 65.5\\
 \multirow{-6}{*}{95} & \multirow{-6}{*}{38.3} &109.0 & 64.9 $\pm$ 1.3 & 57.3 & 60.5 & 65.7 & 66.4 & 65.9\\
 \hline
 & & 69.9 & 63.3 $\pm$ 1.0 & 50.3 & 59.2 & 62.8 & 65.0 & 64.0\\
 & & 79.8 & 66.3 $\pm$ 0.9 & 52.3 & 60.6 & 64.8 & 66.6 & 66.0\\
 & & 89.8 & 67.5 $\pm$ 1.0 & 54.0 & 61.7 & 66.4 & 67.7 & 67.4\\
 & & 99.8 & 68.3 $\pm$ 0.8 & 55.4 & 62.5 & 67.6 & 68.6 & 68.5\\
 \multirow{-5}{*}{100} & \multirow{-5}{*}{40.2} &106.8 & 68.1 $\pm$ 1.2 & 56.2 & 62.9 & 68.3 & 69.1 & 69.1\\
 \hline
 & & 73.4 & 66.8 $\pm$ 1.0 & 55.4 & 61.7 & 65.3 & 67.7 & 66.9\\
 & & 83.8 & 69.6 $\pm$ 1.0 & 57.6 & 63.3 & 67.6 & 69.4 & 69.2\\
 & & 94.3 & 71.0 $\pm$ 1.1 & 59.3 & 64.5 & 69.3 & 70.8 & 70.9\\
 & & 104.8 & 71.5 $\pm$ 0.8 & 60.7 & 65.4 & 70.7 & 71.8 & 72.1\\
 \multirow{-5}{*}{105} & \multirow{-5}{*}{42.2} &112.1 & 71.8 $\pm$ 1.3 & 61.6 & 65.9 & 71.5 & 72.4 & 72.7\\
 \hline
 & & 76.8 & 69.0 $\pm$ 2.0 & 56.9 & 63.9 & 67.5 & 70.1 & 69.6\\
 & & 87.8 & 72.4 $\pm$ 1.2 & 59.2 & 65.6 & 70.0 & 72.0 & 72.0\\
 & & 98.8 & 74.0 $\pm$ 1.2 & 61.1 & 67.0 & 71.9 & 73.5 & 73.9\\
 \multirow{-4}{*}{110} & \multirow{-4}{*}{44.0} &109.8 & 76.1 $\pm$ 1.4 & 62.7 & 68.0 & 73.4 & 74.7 & 75.3\\
 \hline
 & & 64.9 & 68.4 $\pm$ 1.8 & 52.9 & 63.1 & 66.2 & 69.3 & 69.3\\
 & & 69.9 & 71.2 $\pm$ 1.7 & 54.5 & 64.8 & 68.4 & 71.1 & 71.7\\
 \multirow{-3}{*}{130} & \multirow{-3}{*}{50.8} &74.9 & 73.5 $\pm$ 1.7 & 56.0 & 66.2 & 70.4 & 72.7 & 73.8\\
 \hline
 & & 64.9 & 72.5 $\pm$ 1.5 & 52.9 & 63.1 & 66.3 & 69.3 & 69.4\\
 & & 69.9 & 72.2 $\pm$ 1.5 & 54.5 & 64.8 & 68.5 & 71.2 & 71.8\\
 \multirow{-3}{*}{131} & \multirow{-3}{*}{51.2} &74.9 & 73.6 $\pm$ 1.6 & 56.0 & 66.3 & 70.5 & 72.8 & 73.9\\
 \hline
 & & 64.9 & 69.6 $\pm$ 2.6 & 52.9 & 63.2 & 66.4 & 69.4 & 69.6\\
 & & 69.9 & 72.3 $\pm$ 2.1 & 54.6 & 65.0 & 68.6 & 71.3 & 72.1\\
 \multirow{-3}{*}{132} & \multirow{-3}{*}{51.6} &74.9 & 74.7 $\pm$ 1.8 & 56.1 & 66.5 & 70.7 & 73.0 & 74.2\\
 \hline
 & & 55.0 & 64.5 $\pm$ 1.1 & 49.6 & 59.3 & 61.1 & 65.1 & 63.8\\
 & & 61.6 & 67.6 $\pm$ 0.6 & 52.2 & 62.1 & 64.8 & 68.1 & 67.9\\
 & & 64.9 & 69.9 $\pm$ 1.7 & 53.4 & 63.3 & 66.4 & 69.4 & 69.7\\
 & & 68.3 & 71.6 $\pm$ 0.9 & 54.6 & 64.5 & 68.0 & 70.8 & 71.4\\
 & & 69.9 & 72.9 $\pm$ 1.6 & 55.1 & 65.0 & 68.7 & 71.3 & 72.1\\
 & & 74.9 & 75.2 $\pm$ 1.6 & 56.6 & 66.5 & 70.8 & 73.0 & 74.4\\
 \multirow{-7}{*}{133} & \multirow{-7}{*}{52.0} &75.9 & 75.3 $\pm$ 1.0 & 56.9 & 66.8 & 71.2 & 73.4 & 74.8\\
 \hline 
 \\ \\
 & & 64.9 & 69.9 $\pm$ 1.6 & 54.4 & 63.3 & 66.5 & 69.5 & 69.8\\
 & & 69.9 & 72.9 $\pm$ 1.6 & 56.1 & 65.1 & 68.8 & 71.4 & 72.3\\
 \multirow{-3}{*}{134} & \multirow{-3}{*}{52.3} &74.9 & 75.0 $\pm$ 1.6 & 57.7 & 66.6 & 70.9 & 73.1 & 74.5\\
 \hline
 & & 64.9 & 70.1 $\pm$ 1.4 & 57.0 & 63.5 & 66.6 & 69.7 & 70.1\\
 & & 69.9 & 72.8 $\pm$ 1.5 & 59.0 & 65.3 & 69.0 & 71.6 & 72.7\\
 \multirow{-3}{*}{135} & \multirow{-3}{*}{52.9} &74.9 & 75.2 $\pm$ 1.5 & 60.7 & 66.9 & 71.1 & 73.4 & 75.0\\
 \hline
 & & 59.9 & 66.9 $\pm$ 0.9 & 57.2 & 61.6 & 64.1 & 67.6 & 67.4\\
 & & 64.9 & 70.3 $\pm$ 0.4 & 59.5 & 63.6 & 66.8 & 69.8 & 70.3\\
 & & 69.9 & 73.4 $\pm$ 0.5 & 61.6 & 65.4 & 69.2 & 71.8 & 72.9\\
 \multirow{-4}{*}{136} & \multirow{-4}{*}{53.4} &74.9 & 75.3 $\pm$ 0.6 & 63.4 & 67.0 & 71.4 & 73.6 & 75.3\\
 \hline
 & & 64.9 & 70.6 $\pm$ 1.6 & 60.2 & 63.6 & 66.9 & 69.8 & 70.4\\
 & & 69.9 & 72.8 $\pm$ 1.5 & 62.3 & 65.4 & 69.3 & 71.8 & 73.0\\
 \multirow{-3}{*}{137} & \multirow{-3}{*}{53.7} &74.9 & 74.9 $\pm$ 1.6 & 64.2 & 67.1 & 71.5 & 73.6 & 75.4\\
 \hline
 & & 59.9 & 67.3 $\pm$ 1.5 & 58.0 & 61.6 & 64.2 & 67.6 & 67.5\\
 & & 64.9 & 71.2 $\pm$ 1.5 & 60.3 & 63.7 & 66.9 & 69.9 & 70.5\\
 \multirow{-3}{*}{138} & \multirow{-3}{*}{54.1} &69.9 & 73.6 $\pm$ 1.5 & 62.4 & 65.5 & 69.4 & 71.9 & 73.2\\
 \hline
 & & 59.9 & 68.7 $\pm$ 1.5 & 55.5 & 61.7 & 64.2 & 67.7 & 67.5\\
 & & 64.9 & 72.2 $\pm$ 1.4 & 57.7 & 63.7 & 67.0 & 69.9 & 70.6\\
 \multirow{-3}{*}{139} & \multirow{-3}{*}{54.5} &69.9 & 74.6 $\pm$ 1.5 & 59.7 & 65.6 & 69.4 & 72.0 & 73.3\\
 \hline
 & & 57.9 & 65.2 $\pm$ 1.1 & 55.7 & 60.7 & 62.9 & 66.7 & 66.1\\
 & & 64.9 & 68.7 $\pm$ 0.7 & 58.9 & 63.7 & 66.9 & 69.9 & 70.5\\
 \multirow{-3}{*}{140} & \multirow{-3}{*}{54.7} &71.9 & 72.8 $\pm$ 1.0 & 61.7 & 66.2 & 70.3 & 72.7 & 74.3\\
 \hline
 & & 54.9 & 65.1 $\pm$ 0.9 & 48.5 & 59.4 & 61.2 & 65.2 & 64.0\\
 & & 64.9 & 71.7 $\pm$ 0.6 & 52.6 & 63.7 & 67.1 & 69.9 & 70.5\\
 & & 69.9 & 74.4 $\pm$ 0.9 & 54.3 & 65.5 & 69.7 & 71.9 & 73.2\\
 \multirow{-4}{*}{140} & \multirow{-4}{*}{pure 55} &74.9 & 77.8 $\pm$ 1.1 & 55.9 & 67.2 & 71.9 & 73.8 & 75.7\\
 \hline
 & & 59.9 & 68.3 $\pm$ 1.4 & 51.5 & 61.7 & 64.3 & 67.8 & 67.6\\
 & & 64.9 & 70.9 $\pm$ 1.4 & 53.5 & 63.8 & 67.1 & 70.0 & 70.7\\
 \multirow{-3}{*}{141} & \multirow{-3}{*}{55.2} &69.9 & 73.7 $\pm$ 1.5 & 55.3 & 65.7 & 69.6 & 72.1 & 73.5\\
 \hline
 & & 59.9 & 68.6 $\pm$ 1.3 & 50.6 & 61.8 & 64.3 & 67.8 & 67.7\\
 & & 64.9 & 71.5 $\pm$ 1.4 & 52.5 & 63.9 & 67.1 & 70.1 & 70.9\\
 \multirow{-3}{*}{142} & \multirow{-3}{*}{55.7} &69.9 & 73.5 $\pm$ 1.5 & 54.3 & 65.8 & 69.7 & 72.2 & 73.7\\
 \hline
 & & 59.9 & 67.8 $\pm$ 1.3 & 50.3 & 61.8 & 64.2 & 67.8 & 67.7\\
 & & 64.9 & 70.4 $\pm$ 1.4 & 52.2 & 63.9 & 67.1 & 70.1 & 70.9\\
 \multirow{-3}{*}{143} & \multirow{-3}{*}{56.0} &69.9 & 73.3 $\pm$ 1.5 & 53.9 & 65.8 & 69.7 & 72.3 & 73.7\\
 \hline
 & & 59.9 & 67.9 $\pm$ 1.7 & 49.9 & 61.8 & 63.3 & 67.9 & 67.7\\
 \multirow{-2}{*}{144} & \multirow{-2}{*}{pure 55} &64.9 & 73.0 $\pm$ 1.5 & 51.9 & 63.9 & 66.2 & 70.2 & 70.9\\
 \hline
 & & 49.9 & 61.0 $\pm$ 0.5 & 45.9 & 57.1 & 57.3 & 62.6 & 60.0\\
 & & 54.9 & 63.6 $\pm$ 0.8 & 48.1 & 59.5 & 60.9 & 65.4 & 64.1\\
 & & 59.9 & 67.1 $\pm$ 0.5 & 50.1 & 61.8 & 64.2 & 67.9 & 67.7\\
 \multirow{-4}{*}{144} & \multirow{-4}{*}{56.3} &64.9 & 70.2 $\pm$ 0.8 & 52.0 & 63.9 & 67.1 & 70.2 & 70.9\\
 \hline
 & & 54.9 & 63.5 $\pm$ 1.8 & 48.1 & 59.6 & 60.9 & 65.4 & 64.1\\
 & & 59.9 & 67.9 $\pm$ 1.5 & 50.1 & 61.9 & 64.2 & 68.0 & 67.8\\
 \multirow{-3}{*}{145} & \multirow{-3}{*}{56.8} &64.9 & 70.7 $\pm$ 1.5 & 51.9 & 64.1 & 67.1 & 70.3 & 71.0\\
 \hline
 & & 54.9 & 63.6 $\pm$ 1.3 & 48.1 & 59.7 & 61.0 & 65.5 & 64.2\\
 & & 59.9 & 67.9 $\pm$ 1.3 & 50.1 & 62.0 & 64.3 & 68.0 & 67.9\\
 \multirow{-3}{*}{146} & \multirow{-3}{*}{57.3} &64.9 & 70.6 $\pm$ 1.5 & 52.0 & 64.2 & 67.3 & 70.4 & 71.2\\
 \hline
 & & 54.9 & 63.3 $\pm$ 1.4 & 48.0 & 59.7 & 60.8 & 65.5 & 64.1\\
 & & 59.9 & 67.5 $\pm$ 1.6 & 50.0 & 62.0 & 64.2 & 68.0 & 67.8\\
 \multirow{-3}{*}{147} & \multirow{-3}{*}{57.6} &64.9 & 70.7 $\pm$ 1.6 & 51.8 & 64.2 & 67.2 & 70.4 & 71.2\\
 \hline
\end{longtable}

\twocolumn

\onecolumn

\begin{longtable}{c c  c  c c c c c c}
	
	\caption{Stopping power in nickel obtained in this work, compared to the values of SRIM-2013 \cite{ziegler_srim_2010}, DPASS 21.06 \cite{schinner_expanded_2019} and the Knyazheva et al. model \cite{KNYAZHEVA2006}. The last two columns shows the rescaled DPASS values and the fitted  Knyazheva et al.  model respectively. }
	\label{tab:results_Ni}\\
	
	\hline
	\multirow{2}{*}{Mass} & \multirow{2}{*}{<Z>} & \multirow{2}{*}{Energy (MeV)} & \multicolumn{6}{c}{Stopping power (MeV.cm$^2$.mg$^{-1}$)}  \\
	& & & This work & TRIM & DPASS & Knyazheva & Fit DPASS & Fit Knyazheva \\
	\hline
	\endhead
	
	\hline
	\endfoot
	 & & 72.1 & 26.3 $\pm$ 1.0 & 25.0 & 28.7 & 24.8 & 26.9 & 27.0\\
	& & 76.3 & 26.0 $\pm$ 0.8 & 25.5 & 29.1 & 25.3 & 27.5 & 27.5\\
	& & 84.8 & 28.1 $\pm$ 0.5 & 26.4 & 29.8 & 26.1 & 28.4 & 28.5\\
	\multirow{-4}{*}{85} & \multirow{-4}{*}{34.3} &89.0 & 28.5 $\pm$ 0.7 & 26.9 & 30.0 & 26.4 & 28.8 & 28.9\\
	\hline
	& & 76.3 & 27.7 $\pm$ 0.6 & 26.5 & 30.1 & 26.0 & 28.2 & 28.2\\
	& & 80.8 & 28.4 $\pm$ 0.6 & 27.0 & 30.6 & 26.5 & 28.8 & 28.9\\
	& & 89.8 & 30.0 $\pm$ 0.4 & 27.9 & 31.3 & 27.4 & 29.9 & 29.9\\
	\multirow{-4}{*}{90} & \multirow{-4}{*}{36.3} &94.3 & 30.5 $\pm$ 0.6 & 28.2 & 31.6 & 27.8 & 30.3 & 30.4\\
	\hline
	& & 80.6 & 29.4 $\pm$ 0.5 & 27.1 & 31.6 & 27.1 & 29.5 & 29.5\\
	& & 85.3 & 30.2 $\pm$ 0.6 & 27.7 & 32.0 & 27.7 & 30.1 & 30.1\\
	& & 94.8 & 31.6 $\pm$ 0.4 & 28.8 & 32.9 & 28.6 & 31.3 & 31.3\\
	\multirow{-4}{*}{95} & \multirow{-4}{*}{38.3} &99.5 & 32.2 $\pm$ 0.6 & 29.3 & 33.2 & 29.1 & 31.8 & 31.8\\
	\hline
	& & 84.8 & 30.4 $\pm$ 0.5 & 26.7 & 32.9 & 28.2 & 30.7 & 30.6\\
	& & 89.8 & 31.5 $\pm$ 0.6 & 27.3 & 33.5 & 28.8 & 31.4 & 31.4\\
	& & 99.8 & 33.0 $\pm$ 0.4 & 28.5 & 34.4 & 29.8 & 32.6 & 32.6\\
	\multirow{-4}{*}{100} & \multirow{-4}{*}{40.2} &104.8 & 33.7 $\pm$ 0.7 & 29.1 & 34.7 & 30.3 & 33.2 & 33.2\\
	\hline
	& & 89.0 & 31.6 $\pm$ 0.6 & 29.3 & 34.3 & 29.3 & 31.9 & 31.8\\
	& & 94.3 & 32.6 $\pm$ 0.6 & 30.0 & 34.9 & 29.9 & 32.7 & 32.6\\
	& & 104.8 & 34.1 $\pm$ 0.5 & 31.3 & 35.9 & 31.0 & 34.0 & 34.0\\
	\multirow{-4}{*}{105} & \multirow{-4}{*}{42.2} &110.0 & 34.6 $\pm$ 0.7 & 31.9 & 36.3 & 31.5 & 34.6 & 34.6\\
	\hline
	& & 93.3 & 33.1 $\pm$ 0.8 & 30.2 & 35.6 & 30.2 & 33.0 & 32.9\\
	& & 98.8 & 33.8 $\pm$ 0.7 & 30.9 & 36.2 & 30.9 & 33.8 & 33.7\\
	\multirow{-3}{*}{110} & \multirow{-3}{*}{44.0} &109.7 & 35.8 $\pm$ 0.7 & 32.3 & 37.3 & 32.1 & 35.3 & 35.2\\
	\hline
	& & 64.9 & 28.0 $\pm$ 0.6 & 24.5 & 32.2 & 25.5 & 27.1 & 27.1\\
	\multirow{-2}{*}{130} & \multirow{-2}{*}{50.8} &74.8 & 29.6 $\pm$ 0.7 & 26.4 & 34.3 & 27.7 & 29.6 & 29.6\\
	\hline
	& & 64.9 & 28.3 $\pm$ 0.5 & 24.5 & 32.2 & 25.5 & 27.0 & 27.0\\
	& & 69.8 & 29.5 $\pm$ 0.5 & 25.5 & 33.3 & 26.7 & 28.4 & 28.4\\
	\multirow{-3}{*}{131} & \multirow{-3}{*}{51.2} &74.8 & 30.5 $\pm$ 0.6 & 26.4 & 34.3 & 27.7 & 29.6 & 29.6\\
	\hline
	& & 55.0 & 24.1 $\pm$ 0.4 & 22.5 & 29.9 & 23.0 & 24.1 & 24.2\\
	& & 61.6 & 26.3 $\pm$ 0.3 & 24.0 & 31.6 & 24.7 & 26.1 & 26.1\\
	& & 68.2 & 27.9 $\pm$ 0.3 & 25.4 & 33.1 & 26.3 & 27.9 & 27.9\\
	\multirow{-4}{*}{133} & \multirow{-4}{*}{52.0} &75.8 & 29.4 $\pm$ 0.4 & 26.9 & 34.7 & 28.0 & 29.9 & 29.8\\
	\hline
	& & 59.9 & 26.1 $\pm$ 0.5 & 26.1 & 31.4 & 24.3 & 25.6 & 25.7\\
	& & 64.9 & 27.2 $\pm$ 0.2 & 27.4 & 32.6 & 25.5 & 27.1 & 27.1\\
	& & 69.8 & 28.5 $\pm$ 0.2 & 28.6 & 33.7 & 26.7 & 28.4 & 28.4\\
	\multirow{-4}{*}{136} & \multirow{-4}{*}{53.4} &74.8 & 29.4 $\pm$ 0.2 & 29.8 & 34.8 & 27.8 & 29.7 & 29.7\\
	\hline
	& & 57.9 & 24.9 $\pm$ 0.5 & 25.3 & 30.9 & 23.7 & 24.8 & 24.9\\
	& & 64.9 & 26.6 $\pm$ 0.3 & 27.1 & 32.7 & 25.5 & 26.9 & 26.9\\
	& & 71.8 & 28.4 $\pm$ 0.4 & 28.7 & 34.3 & 27.1 & 28.8 & 28.8\\
	\multirow{-4}{*}{140} & \multirow{-4}{*}{54.7} &79.8 & 29.0 $\pm$ 1.0 & 30.4 & 35.9 & 28.8 & 30.9 & 30.8\\
	\hline
	& & 60.8 & 25.4 $\pm$ 0.4 & 22.9 & 31.9 & 24.3 & 25.6 & 25.7\\
	\multirow{-2}{*}{147} & \multirow{-2}{*}{57.6} &68.1 & 25.8 $\pm$ 0.5 & 24.4 & 33.8 & 26.2 & 27.7 & 27.7\\
	\hline
	& & 49.9 & 21.0 $\pm$ 1.4 & 20.3 & 28.9 & 21.2 & 22.0 & 22.4\\
	& & 54.9 & 24.7 $\pm$ 0.7 & 21.5 & 30.3 & 22.6 & 23.7 & 23.9\\
	\multirow{-3}{*}{149} & \multirow{-3}{*}{58.3} &59.9 & 26.0 $\pm$ 0.5 & 22.6 & 31.7 & 24.0 & 25.2 & 25.3\\
	\hline
	
\end{longtable}

\twocolumn

\end{document}